\documentclass[twocolumn,preprintnumbers,floats,prd,amssymb,floatfix,nofootinbib,balancelastpage,superscriptaddress,amsmath]{revtex4-1}

\pdfoutput=1
\usepackage[utf8]{inputenc}
\usepackage{amssymb}
\usepackage{amsmath}
\usepackage{amsfonts}
\usepackage{graphicx}
\usepackage{color}
\usepackage{xspace}
\usepackage{comment}
\usepackage{hyperref}
\usepackage[normalem]{ulem}
\usepackage[section]{placeins}
\usepackage{afterpage}
\usepackage{slashed}
\usepackage{enumerate}
\usepackage{enumitem}
\usepackage{caption}
\usepackage{subfigure}
\usepackage{float}
\usepackage{ulem}
\usepackage{verbatim}
\usepackage{multirow,rotating}
\usepackage[dvipsnames]{xcolor}

\definecolor{dcolour}{rgb}{.5, .5, .5}
\def\gsim{\raise0.3ex\hbox{$\;>$\kern-0.75em\raise-1.1ex\hbox{$\sim\;$}}}
\def\lsim{\raise0.3ex\hbox{$\;<$\kern-0.75em\raise-1.1ex\hbox{$\sim\;$}}}
\def\gsim{\raise0.3ex\hbox{$\;>$\kern-0.75em\raise-1.1ex\hbox{$\sim\;$}}}
\def\lsim{\raise0.3ex\hbox{$\;<$\kern-0.75em\raise-1.1ex\hbox{$\sim\;$}}}

\newcommand{\ba}[1]{\begin{eqnarray} \label{(#1)}}
\newcommand{\ea}{\end{eqnarray}}

\newcommand{\iab}{\rm ab^{-1}}
\newcommand{\ifb}{\rm fb^{-1}}
\newcommand{\mltp}{{\mkern-2mu\times\mkern-2mu}}

\newcommand{\met}{\slashed{E}_T}

\begin{document}

\title{Search for heavy Majorana neutrinos at electron-proton colliders}

\author{Haiyong Gu}
\email{haiyong.gu@whut.edu.cn}
\affiliation{Department of Physics, School of Science, Wuhan University of Technology, \\430070 Wuhan, Hubei, China }

\author{Kechen Wang}
\email{kechen.wang@whut.edu.cn (Corresponding author)}
\affiliation{Department of Physics, School of Science, Wuhan University of Technology, \\430070 Wuhan, Hubei, China }


\begin{abstract}
We develop the search strategy for a heavy Majorana neutrino via the lepton number violation signal process $p\, e^- \to \mu^+ jjj$ at future electron-proton colliders.
The signal and dominant standard model background events are generated with the fast detector simulation.
We apply the pre-selection criteria and perform the multi-variate analysis based on machine-learning to reject the background.
Distributions of representative kinematic observables are presented for both signal and background processes and effects on final limits are compared by inputting two different set of observables when performing multi-variate analysis.
The 2- and 5-$\sigma$ limits on the mixing parameter $|V_{\ell N}|^2$ are predicted for the heavy neutrino mass $m_N$ in the range of 10$-$1000 GeV.
At the LHeC (FCC-eh) with an electron beam energy of 60 GeV, a proton beam energy of 
7 (50) TeV and an integrated luminosity of 1 (3) ab$^{-1}$, the mixing parameter $|V_{\ell N}|^2$ can be constrained to be below $\sim 3.0~(1.0) \times 10^{-6}$ 
for $m_N$ around $\mathcal{O}(100)$ GeV at 2-$\sigma$ level.
The limits are much stronger than the current experiment limits at the LHC for $m_N$ above 30 GeV.
The positron signal final state and the effect of long-lived cases of heavy neutrinos are also checked and commented.
\end{abstract}


\vskip10mm

\maketitle
\flushbottom


\section{Introduction}
\label{sec:intro}

The neutrino oscillation experiments~\cite{Super-Kamiokande:1998kpq,MINOS:2006foh,MINOS:2011amj,PhysRevLett.108.131801,Ling:2013fta,Kim:2013sza} have proved that neutrinos in standard model (SM) have very tiny masses. 
Because of the lack of right-handed neutrinos in the SM, the Dirac mass terms cannot be formed as other fermions in the Lagrangian and the SM needs to be extended to explain their non-zero masses.
One important solution is the seesaw mechanism~\cite{FRITZSCH1975256,Minkowski:1977sc,Yanagida:1979as,Sawada:1979dis,Mohapatra:1979ia,Glashow:1979nm,GellMann:1980vs,Keung:1983uu,Foot:1988aq,Mohapatra:1986aw,MAGG198061}, where new gauge singlet right-handed neutrinos $N_{R}$ are introduced and masses of active neutrinos are generated by mixing SM left-handed neutrinos $\nu_L$ with right-handed neutrinos $N_{R}$, resulting in heavy mass eigenstates $N$ that have small SM $\nu_L$ components.
Therefore, searches for heavy neutrinos are crucial to verify the seesaw mechanism and explore the origin of neutrino masses.

At colliders, such heavy neutrinos are usually also called heavy neutral leptons and are extensively searched relying on their effective couplings to SM gauge bosons via their mixing with SM neutrinos.
In theory, the production cross section, decay width, and lifetime of $N$ depend on its mass $m_N$ and the parameter $|V_{\ell N}|^2$ which is related to the matrix element describing the mixing of $N$ with the SM neutrino of flavor $\ell$.
Therefore, limits for such searches are usually given in the plane of the mixing parameter $|V_{\ell N}|^2$ vs. the heavy neutrino mass $m_N$.
Summaries of collider searches of heavy neutrinos
can be found in Refs.~\cite{Atre:2009rg,Deppisch:2015qwa,Das:2015toa,Cai:2017mow,Das:2017rsu,Bolton:2019pcu} and references therein.

Refs.~\cite{CMS:2018iaf, CMS:2018jxx, CMS:2021lzm, ATLAS:2019kpx, LHCb:2020wxx, NA62:2020mcv, Belle:2013ytx, T2K:2019jwa}
are recent experimental studies on heavy neutrino searches.
Among them, the CMS collaboration has analyzed the data with center-of-mass energy $\sqrt{s} =$ 13 TeV and an integrated luminosity of 35.9 $\ifb$~\cite{CMS:2018iaf,CMS:2018jxx} and 137 $\ifb$\cite{CMS:2021lzm}.
Ref.~\cite{CMS:2018iaf} searched for a heavy Majorana neutrino in the trilepton signal process $p p \to W^{(*)} \to \ell\, N \to \ell\, (\ell W^{(*)}) \to \ell\, (\ell \ell \nu)$
and limits are applied on both $|V_{e N}|^2$ and $|V_{\mu N}|^2$ for the heavy neutrino mass range between 1 GeV and 1.2 TeV.
Ref.~\cite{CMS:2018jxx} searched for a heavy Majorana neutrino in the same-sign dilepton channel $ W^{(*)} \to \ell^{\pm}\, N \to \ell^{\pm}\, (\ell^\pm W^{(*)}) \to \ell^\pm\, (\ell^\pm j j)$.
The upper limits are set on $|V_{e N}|^2$, $|V_{\mu N}|^2$, and $|V_{eN }V_{\mu N }^*|^2/(|V_{eN }|^2+|V_{\mu N }|^2)$ for $N$ masses between 20 and 1600 GeV.
Ref.~\cite{CMS:2021lzm} searched for a long-lived heavy neutrino of either Majorana or Dirac type in the signal process $p p \to W \to \ell\, N \to \ell\, (\ell W^* / \nu Z^*) \to \ell \ell \ell \nu$. The final state consists of three leptons among which two leptons form a displaced vertex with respect to the primary proton-proton collision vertex and the third lepton emerges from the primary vertex. Limits are applied on $|V_{e N}|^2$ and $|V_{\mu N}|^2$ in the mass range between 1 and 15 GeV.

In Ref.~\cite{ATLAS:2019kpx}, a heavy Majorana neutrino was investigated by the ATLAS collaboration in the similar signal process to that in Ref.~\cite{CMS:2018iaf}. 
Data at $\sqrt{s} =$ 13 TeV and integrated luminosity of 36.1 (32.9) $\iab$ are analysed for the prompt (displaced) leptonic decay case. Constraints on $|V_{e N}|^2$ and $|V_{\mu N}|^2$ are set for heavy neutrino mass in the range of 4.5 to 50 GeV .
The LHCb collaboration searched for a heavy Majorana neutrino in the signal process $W^+ \to \mu^+\, N \to \mu^+\, (\mu^{\pm} j j)$~\cite{LHCb:2020wxx}. Data corresponding to an integrated luminosity of 3.0 $\ifb$ and center-of-mass energies of 7 and 8 TeV are analysed and upper limits on $|V_{\mu N}|^2$ are set to be $\sim 10^{-3}\,\, (10^{-4})$ in the mass range from 5 to 50 GeV for the lepton number conserving (violating) case.

The NA62 collaboration searched for $N$ produced from $K^+ \to e^+ N$ decays and placed the upper limit on $|V_{e N}|^2 \sim 10^{-9}$ in the mass range 144$-$462 MeV~\cite{NA62:2020mcv}. 
Heavy neutrinos from B-meson decays are investigated by the Belle collaboration and upper limits are set on $|V_{e N}|^2$, $|V_{\mu N}|^2$, $|V_{e N}|\,|V_{\mu N}|$ in the mass range 0.5$-$5.0 GeV~\cite{Belle:2013ytx}.
The T2K Collaboration searched for heavy neutrinos from kaon decays and constrained mixing parameters $|V_{\ell N}|^2$ with $\ell = e, \mu, \tau$ for $m_N$ between 140 and 493 MeV~\cite{T2K:2019jwa}.
Moreover, for $N$ produced from meson decays, the SHiP collaboration's search prospect for long-lived neutrinos predicts strong sensitivity 
on $|V_{\ell N}|^2$ with $\ell = e, \mu, \tau$ flavors for $m_N$ in the range $0.1-5.8$ GeV~\cite{SHiP:2018xqw}.
The combination of electroweak precision observables and lepton flavor violating decays can also set  constraints indirectly on mixing parameters  $|V_{\ell N}|^2$, especially when $m_N$ is larger than 80 GeV~\cite{Chrzaszcz:2019inj}.

In this article, we develop the search strategy for a heavy Majorana neutrino via the lepton number violation (LNV) signal process of $p\, e^- \to \mu^+ jjj$ at future electron-proton colliders, the Large Hadron electron Collider (LHeC)~\cite{Klein:2009qt,LHeCStudyGroup:2012zhm,Bruening:2013bga,Klein:2016uwv,LHeC:2020van,Holzer:2021dxw}
and the electron-hadron mode of the Future Circular Collider (FCC-eh)~\cite{Zimmermann:2014qxa,Klein:2016uwv,TOMAS2016149,FCC:2018byv,Holzer:2021dxw}.
We consider the LHeC (FCC-eh) running with an electron beam energy of 60 GeV and a proton beam energy of 
7 (50) TeV, which corresponds to $\sqrt{s} =$ 1.3 (3.5) TeV.
The integrated luminosities are assumed to be 1 and 3 $\iab$ at the LHeC and FCC-eh, respectively.

Compared with the proton colliders, such as the high-luminosity LHC (HL-LHC), the centre-of-mass energy of $ep$ colliders is lower. However, due to lack of gluon-exchange diagrams, the SM QCD backgrounds, which are dominant at $pp$ colliders, are much smaller at $ep$ colliders. Besides, the number of additional interactions in the same event (pileup) is negligible at $ep$ colliders, while it is expected to be very large at the HL-LHC. 
Furthermore, heavy neutrinos with mass above 100 GeV can still be produced on-shell from the $t$-channel exchange of $W$-boson at $ep$ colliders, while at the HL-LHC such heavy neutrinos are produced from off-shell $W$- or $Z$-boson processes with limited cross section.
Therefore, future $e p$ colliders could be complementary to the $p p$ collider when searching for beyond standard model (BSM) physics scenarios, particularly for heavy neutrinos.

Ref.~\cite{Azuelos:2021ese} and references therein have reviewed BSM physics searches at the LHeC and FCC-eh, while phenomenology studies on heavy neutrino searches at $e p$ colliders can be found in Refs.~\cite{BUCHMULLER1991465,Buchmuller:1991tu,Buchmuller:1992wm,Ingelman:1993ve,Liang:2010gm,Blaksley:2011ey,Duarte:2014zea,Mondal:2016kof,Antusch:2016ejd,Lindner:2016lxq,Li:2018wut,Das:2018usr,Antusch:2019eiz,Cottin:2021tfo}.
Among them, Ref.~\cite{Duarte:2014zea} investigated heavy Majorana neutrinos produced in an effective Lagrangian approach at the LHeC.
For the LHeC, they considered a 7 TeV proton beam colliding with an electron beam of two energies: $E_e = $ 50 and 150 GeV.
The events are simulated at the parton level and kinematic cuts are applied to reduce the background.
Limits are placed on the neutrino mass and the effective coupling.
Ref.~\cite{Mondal:2016kof} explored heavy neutrinos at the LHeC in the context of an inverse-seesaw model.
The production cross section of various signals of $N j$, $N j W^-$ and $e^- j W^-$ are calculated with and without $80\%$ left-polarized electron beam.
The events are simulated at the parton level and kinematic cuts are applied to reduce the background.
The required integrated luminosities are estimated to achieve a 3-$\sigma$ statistical significance for two different heavy neutrino masses of 150 and 400 GeV.
Ref.~\cite{Li:2018wut} searched for heavy Majorana neutrino via the signal processes $p\,e^- \to e^- \mu^{\pm} \mu^{\pm} + X$ and $p\,e^- \to \nu_e \mu^- \mu^{\pm} + X$.
The events are simulated including detector smearing effects and kinematic cuts are applied to reduce the background.
The required integrated luminosities are estimated for heavy neutrino masses in the range of 100 and 1000 GeV.
Ref.~\cite{Das:2018usr} investigated a heavy neutrino via the signal process $p\,e^- \to j N \to j (e^\pm W^\mp) \to j (e^\pm J)$, where $J$ is a fat-jet from a highly boosted $W-$boson.
The events are simulated at the parton level and passed through selection cuts to reduce the background.
Bounds on $|V_{e N}|^2$ are placed for $m_N$ in the range of 400 to 1000 GeV.
Ref.~\cite{Antusch:2019eiz} probed heavy neutrinos via the lepton flavor violating signal process $p\,e^- \to j N \to j (\mu^- W^+) \to \mu^- + 3 j$ at the LHeC and FCC-eh. Background processes include $j e^- V V$ and $j \nu_e V V$ where $V = Z, W^\pm$ and $V V \to (j j) (\mu^- \mu^+ / \mu^- \bar{\nu}_\mu)$.
Limits on the mixing parameters $|\theta_e \theta_\mu|$ are placed for the heavy neutrino mass in the range from 100 to 1000 GeV.

We note that this work is different from all previous phenomenology studies due to the combination of following aspects: 
(i) we consider the LNV signal process $p\,e^- \to \mu^+ + jjj$ assuming $|V_{\ell N}|^2 = |V_{e N}|^2 = |V_{\mu N}|^2$ in the context of a simplified Type-I seesaw model;
(ii) SM background includes four inclusive processes listed in Table~\ref{tab:crsc}; 
(iii) for both the signal and background event simulation, we utilize the program chain including the event generator, parton shower, hadronization and detector effects;
(iv) for the LHeC (FCC-eh), we consider a 60 GeV electron beam colliding with a proton beam of 7 (50) TeV energy and an integrated luminosity of 1 (3) $\iab$; 
(v) we apply pre-selection criteria and perform multi-variate analysis based on machine-learning to reject the background; 
(vi) The 2- and 5-$\sigma$ limits on the mixing parameter $|V_{\ell N}|^2$ are predicted for the heavy neutrino mass in the range 10$-$1000 GeV.

The article is organised as follows. 
Sec.~\ref{sec:sig} presents the data simulation and the cross section of the signal process.
Sec.~\ref{sec:bkg} describes the SM background processes and our search strategy.
The analysis details and limits on the mixing parameters $|V_{\ell N}|^2$ at both the LHeC and FCC-eh are shown in Sec.~\ref{sec:result}.
We summarize our study, and comment on the effect of long-lived cases of heavy neutrinos and the positron signal final state in Sec.~\ref{sec:sum}.

\section{The LNV signal}
\label{sec:sig}

To simplify the analyses, we consider the Type-I seesaw model and assume that there is only one generation of heavy neutrinos $N$ which mixes with active neutrinos of electron and muon flavours with the same mixing parameters, i.e. $|V_{\ell N}|^2 = |V_{e N}|^2 = |V_{\mu N}|^2$.
We also assume that $N$ decays promptly in this study.
As shown in Fig.~\ref{fig:Feynman}, the heavy Majorana neutrino $N$ can be produced via the $t$-channel exchange of $W$-boson at $ep$ colliders, and finally decay into $\mu^+$ plus three jets.
The lepton number of this process changes from $+1$ to $-1$, so it violates the conservation of lepton number.

\begin{figure}[h]
\centering
\includegraphics[width=5cm,height=3.5cm]{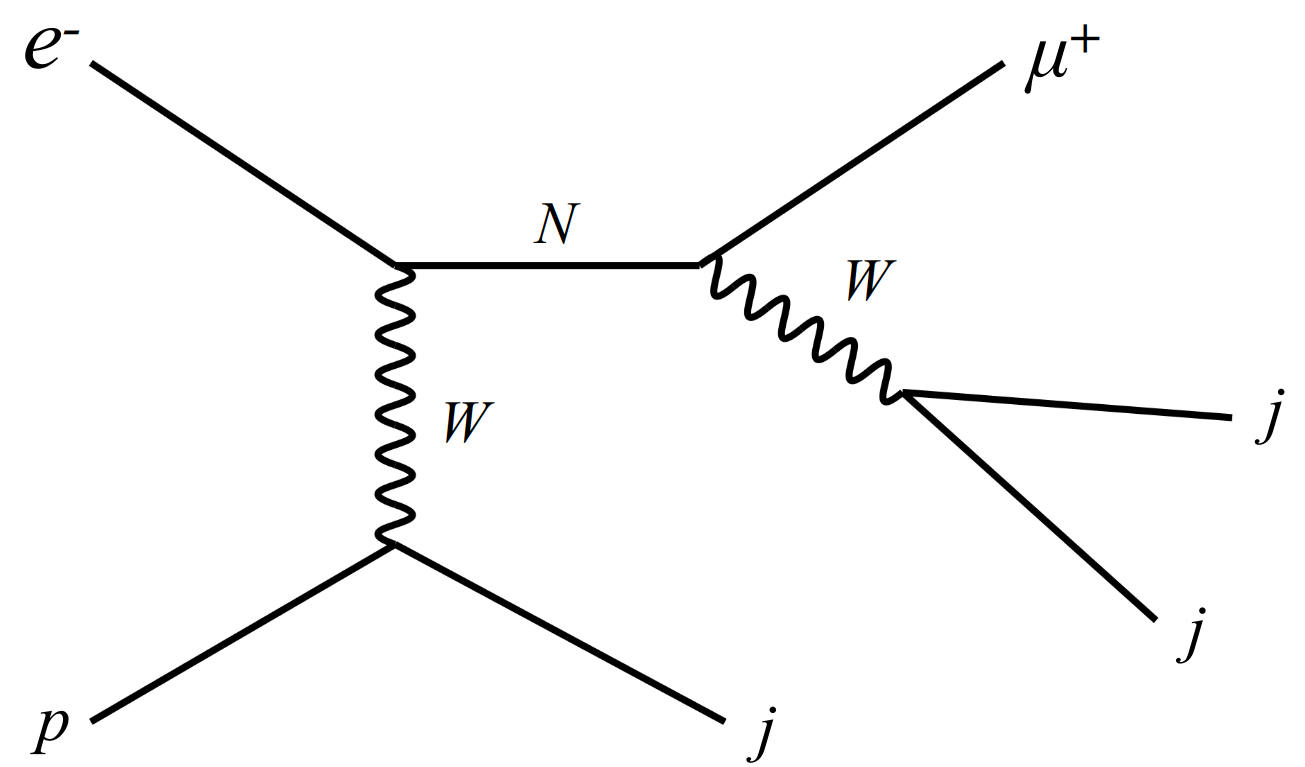}  
\caption{
The production process of the LNV signal via a Majorana heavy neutrino $N$ at $ep$ colliders.
}
\label{fig:Feynman}
\end{figure}

For the data simulation, we implement the Universal FeynRules Output model file~\cite{Alva:2014gxa,Degrande:2016aje} which extends the SM with additional heavy neutrinos interacting with active neutrinos, into the MadGraph5~\cite{Alwall:2014hca} to generate the signal events.
Similar to our previous work~\cite{Azuelos:2019bwg}, the Pythia6~\cite{Sjostrand:2006za} program is modified to perform the parton showering and hadronization for $ep$ colliders, while the configuration card files~\cite{Delphes_cards} for the LHeC and FCC-eh detectors are implemented to the Delphes program~\cite{deFavereau:2013fsa} to complete the detector simulation.

\begin{figure}[]
\centering
\includegraphics[width=8cm,height=6cm]{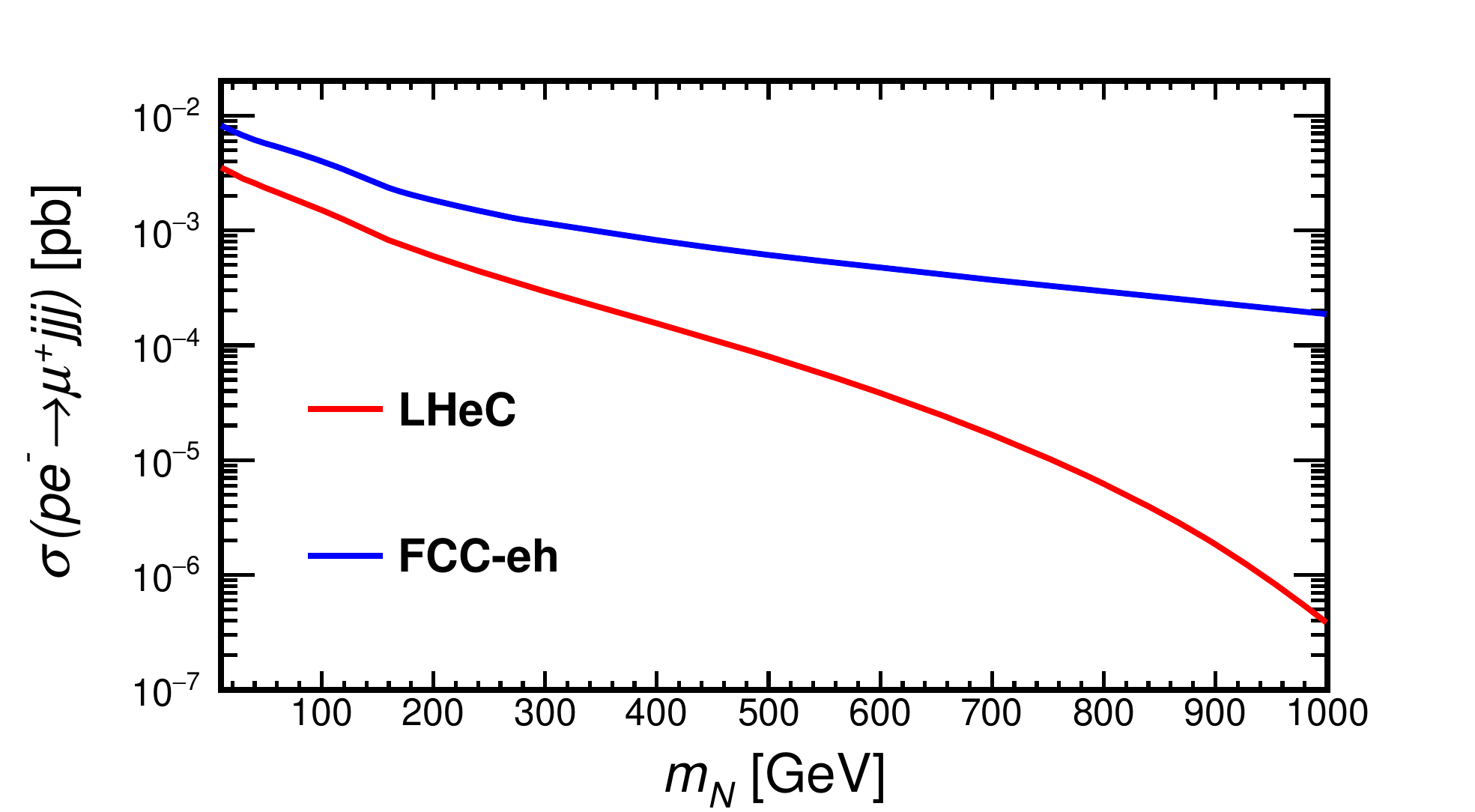}  
\caption{
The production cross section of the LNV signal $p\, e^- \to \mu^+ jjj$ via the heavy Majorana neutrino $N$ for varying heavy neutrino masses $m_N$ at the LHeC and FCC-eh with $|V_{\ell N}|^2 = 10^{-4}$.
}
\label{fig:crs}
\end{figure}

To maintain consistency throughout this study, the production cross sections calculated by MadGraph5 are used to estimate the number of events for both signal and background processes.
In Fig.~\ref{fig:crs}, we plot the  cross sections of the LNV signal $p\, e^- \to \mu^+ jjj$ via the heavy Majorana neutrino $N$ as a function of the heavy neutrino mass $m_N$ at the LHeC and FCC-eh, where the mixing parameter $|V_{\ell N}|^2$ is fixed to be $10^{-4}$.

For large $m_N$, cross sections at the LHeC decrease much faster than those at the FCC-eh. 
This behaviour can be understood from the parton distribution function (PDF) of the proton. 
In this study, heavy neutrinos are produced from the $t$-channel $W$-boson exchange process $q\,e^-  \to j N$.
In order to produce heavy neutrinos with large mass, momenta of incoming quarks need to be large enough, so that the centre of mass energy $\sqrt{s}= 2\sqrt{ E_e E_q}$ is larger than $m_N$.
Considering a quark carries a fraction $x$ of the longitudinal momentum of a proton, only quarks with $x  \gtrsim m_N^2 / (4 E_e E_p)$ can contribute to the signal production cross section.
Since the parton density function $f(x)$ of the quark decreases rapidly for large $x$ values~\cite{Placakyte:2011az}, 
when $m_N$ is very large, the number of quarks which satisfies the condition  $x  \gtrsim m_N^2 / (4 E_e E_p)$ becomes tiny.
This leads to the rapid decrease in the production cross section for large $m_N$.

\section{background and search strategy}
\label{sec:bkg}

Since the signal process $p\, e^- \to \mu^+ jjj$ violates the lepton number conservation explicitly, it has little SM background in theory.
Considering the signal final state contains one positive charged muon plus multi-jets, 
there are mainly four SM background processes, which we label as ``B1 - B4" in this article.
We list their production cross sections in Table~\ref{tab:crsc} .
They can contribute to the background when the final state $e^-$ and/or $\mu ^-$ are undetected. Among these four background processes, $\mu^+ \mu^- e^- j j j$ and $ \mu^+ \nu_\mu e^- j j j$ have large cross sections and are more difficult to eliminate. 
We note that the considered four background processes have both QED and QCD interactions.
Processes with pure QED interactions are checked and the sum of their cross sections are found to be only a factor of about 1/90 (1/45) of the considered background at the LHeC (FCC-eh). Because they are much smaller and our computing resources are limited, we do not include the background processes with only QED interactions.

\begin{table}[h]
\centering
\begin{tabular}{cccc}
\hline 
\hline
& $\sigma$ [pb] & LHeC  & FCC-eh  \\ 	
\hline 
B1 & $\mu^+ \mu^- e^- j j j$& 0.58 & 2.1 \\ 
B2 & $\mu^+ \mu^- \nu_e j j j$& $8.6\mltp 10^{-2}$  &  0.39 \\ 
B3 & $\mu^+ \nu_\mu  e^- j j j$&0.28  & 1.6 \\ 
B4 & $\mu^+ \nu_\mu \nu_e j j j$& $8.1 \mltp 10^{-6}$  & $9.3 \mltp 10^{-5}$  \\ 
\hline 
\hline
\end{tabular} 
\caption{
The production cross sections of dominant background processes at the LHeC and FCC-eh. 
}
\label{tab:crsc}
\end{table}

The final state muon can also come from tau decays.
However, it will not contribute too much to the background because of the following two reasons: (i) the small tau to muon branching ratio; (ii) the leptonic decay of taus produce neutrinos, resulting in large missing energy in the final state, and cannot pass our missing energy pre-selection cut.
We checked four background processes $\tau^+ \tau^- e^- j j j$, $\tau^+ \tau^- \nu_e j j j$, $\tau^+ \nu_\tau e^- j j j$, and $\tau^+ \nu_\tau  \nu_e j j j$ at the LHeC (FCC-eh) and found that after pre-selection the event rate of total background increased only by a factor of $6\%$ ($11\%$).
After performing the full analysis described below, we find that after adding the tau background, the limit on $|V_{\ell N}|^2$ changed from $3.6\,\, (1.10) \times10^{-6}$ to $3.8\,\, (1.12) \times10^{-6}$ for the benchmark $m_N$ = 120 GeV.
Since the effects on the limits are very small, we did not add the tau background in this study.

We apply the following pre-selection cuts to select the signal and reject the background events at the first stage.
(i) Exactly one muon with positive charge, i.e. $N(\mu^+) = 1$ and transverse momentum $p_T(\mu) >$ 5 GeV; 
events with final state electron(s) or tau(s) are vetoed.
(ii) All jets are sorted in descending order according to their transverse momenta and we require at least three jets, i.e. $N(j)\geq 3$; for the $p_T$ thresholds of jets, when heavy neutrino mass is below 80 GeV, the $p_T$ of the first three leading jets are required to be greater than 10 GeV, while when masses are above 80 GeV, we require the first two leading jets to have $p_T$ greater than 20 GeV and the third one to have $p_T$ greater than 10 GeV.
(iii) Since both the final state neutrinos and the missing of leptons contribute to the missing energy, the background has much larger missing energy compared with the signal and a pre-selection of missing energy $\met < 10 $ GeV is applied to reject the background. 

For the signal data simulation, we vary the heavy neutrino mass $m_N$ from 10 to 1000 GeV and generate 0.3 million signal events at the LHeC and FCC-eh for each $m_N$.
Due to limited computational resources, we are not able to generate huge number of events for every background processes.
The number of simulated events for each background process is determined according to its importance in reducing the statistical uncertainty on final limits.
For the background, we generate 2.1 (2.0) million $\mu^+ \mu^- e^- j j j$, 10.5 (6.0) million $\mu^+ \mu^- \nu_e j j j$, 27.4 (24.6) million $\mu^+ \nu_\mu e^- j j j$ and 6.0 (6.4) million  $\mu^+ \nu_\mu  \nu_e j j j$ events at the LHeC (FCC-eh), respectively.
In Table~\ref{tab:cutmn120}, we show the number of events for the signal with benchmark $m_N=120$ GeV and four background processes after applying the pre-selection cuts (i)-(iii) sequentially described above.  

\begin{table}[h]
\centering
\begin{tabular}{ccccccc}
\hline
\hline
& & signal & B1 & B2 & B3 & B4  \\
\hline
\multirow{4}{*}{LHeC} 
& initial &$1.2\mltp10^{3}$ & $5.8\mltp10^{5}$ & $8.6\mltp10^{4}$ & $2.8\mltp10^{5}$ & 8.1  \\ 
& (i) & $1.1\mltp10^{3}$  & $2.6\mltp10^{3}$ & $4.0\mltp10^{3}$  & $1.9\mltp10^{4}$  & 6.2  \\ 
& (ii) & 853 &  799 & $1.3\mltp10^{3}$  & $6.5\mltp10^{3}$ & 4.4  \\ 
& (iii) & 702  & 699  & 9.3  & 154  &  0.1 \\ 
\hline
\multirow{4}{*}{FCC-eh}
&	initial & $1.0\mltp10^{4}$ &  $6.2\mltp10^{6}$ &  $ 1.2\mltp10^{6} $ & $4.9\mltp10^{6} $  &  278 \\ 
& (i) & $8.8\mltp10^{3}$  &$1.3\mltp10^{4}$  &  $6.1\mltp10^{4}$ & $2.8\mltp10^{5}$  &  118 \\ 
& (ii) & $7.2\mltp10^{3}$  & $4.1\mltp10^{3}$  &  $2.3\mltp10^{4}$ &  $1.2\mltp10^{5}$ & 99  \\ 
& (iii) &  $5.5\mltp10^{3}$ & $2.8\mltp10^{3}$    & 125  &  $3.2\mltp10^{3}$ & 1.2  \\ 
\hline
\hline
\end{tabular}
\caption{
The number of events for the signal with benchmark $m_N$ = 120 GeV and four background processes after applying pre-selection cuts (i)-(iii) sequentially. The numbers correspond to the LHeC and FCC-eh with 1 and 3 $\iab$ integrated luminosity, respectively.
}
\label{tab:cutmn120}
\end{table}
To further reject the background, we input the following nineteen observables into the TMVA~\cite{Hocker:2007ht} package to perform the multivariate analysis (MVA).
\begin{enumerate}[label*=\Alph*.]
\item The four-momenta of the final state muon:
$E(\mu)$, $p_{x}(\mu)$, $p_{y}(\mu)$, $p_{z}(\mu)$.
\item The number of jets $N(j)$ and the four-momenta of the first three leading jets:
$E(j_{1})$, $p_{x}(j_{1})$, $p_{y}(j_{1})$, $p_{z}(j_{1})$;
$E(j_{2})$, $p_{x}(j_{2})$, $p_{y}(j_{2})$, $p_{z}(j_{2})$;
$E(j_{3})$, $p_{x}(j_{3})$, $p_{y}(j_{3})$, $p_{z}(j_{3})$.
\item The magnitude and the azimuthal angle of the missing transverse momentum:
$\met$, $\phi(\met)$; 
\end{enumerate}
\begin{figure}[h]
\centering
\includegraphics[width=4cm,height=3cm]{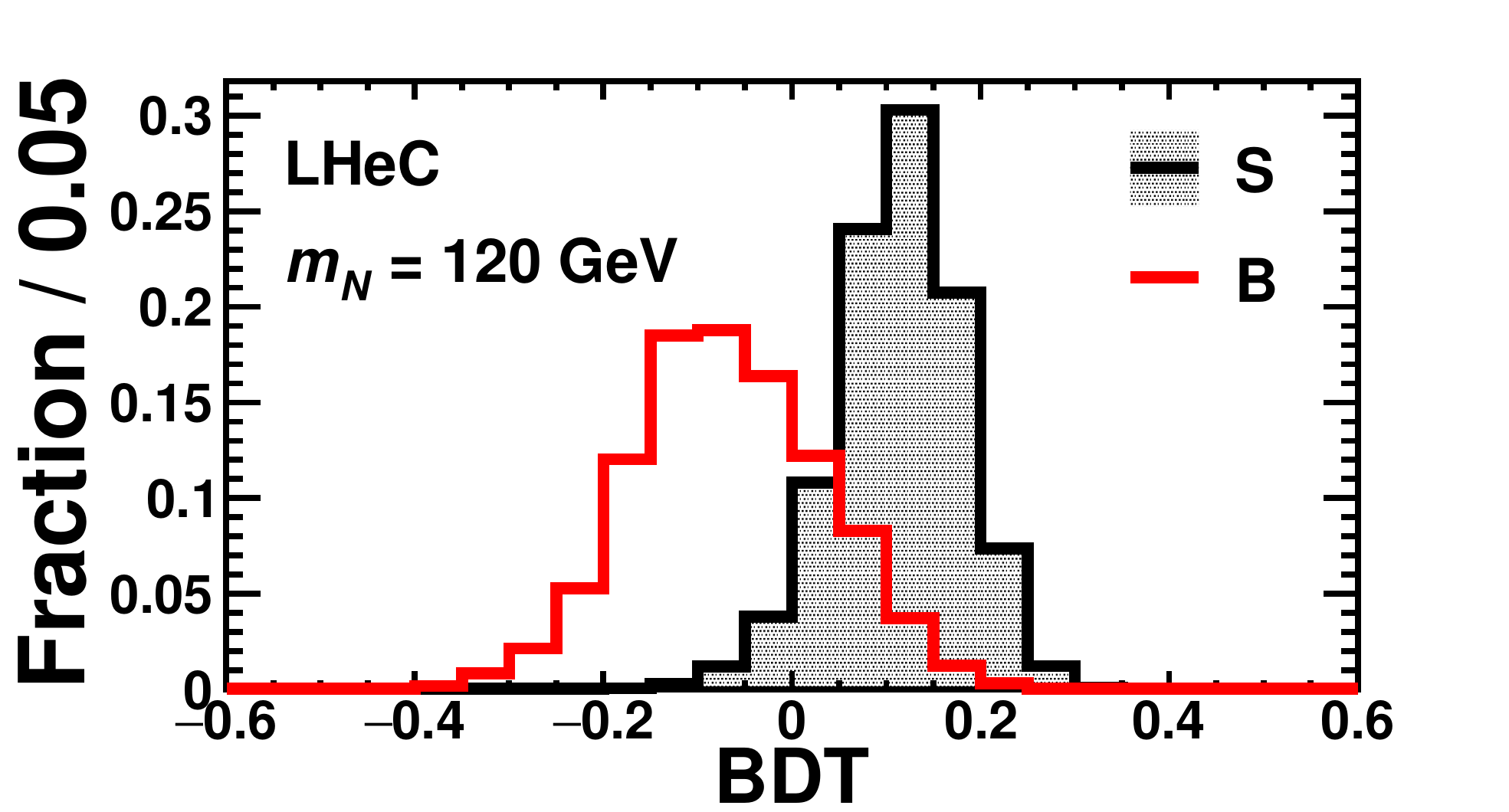}
\includegraphics[width=4cm,height=3cm]{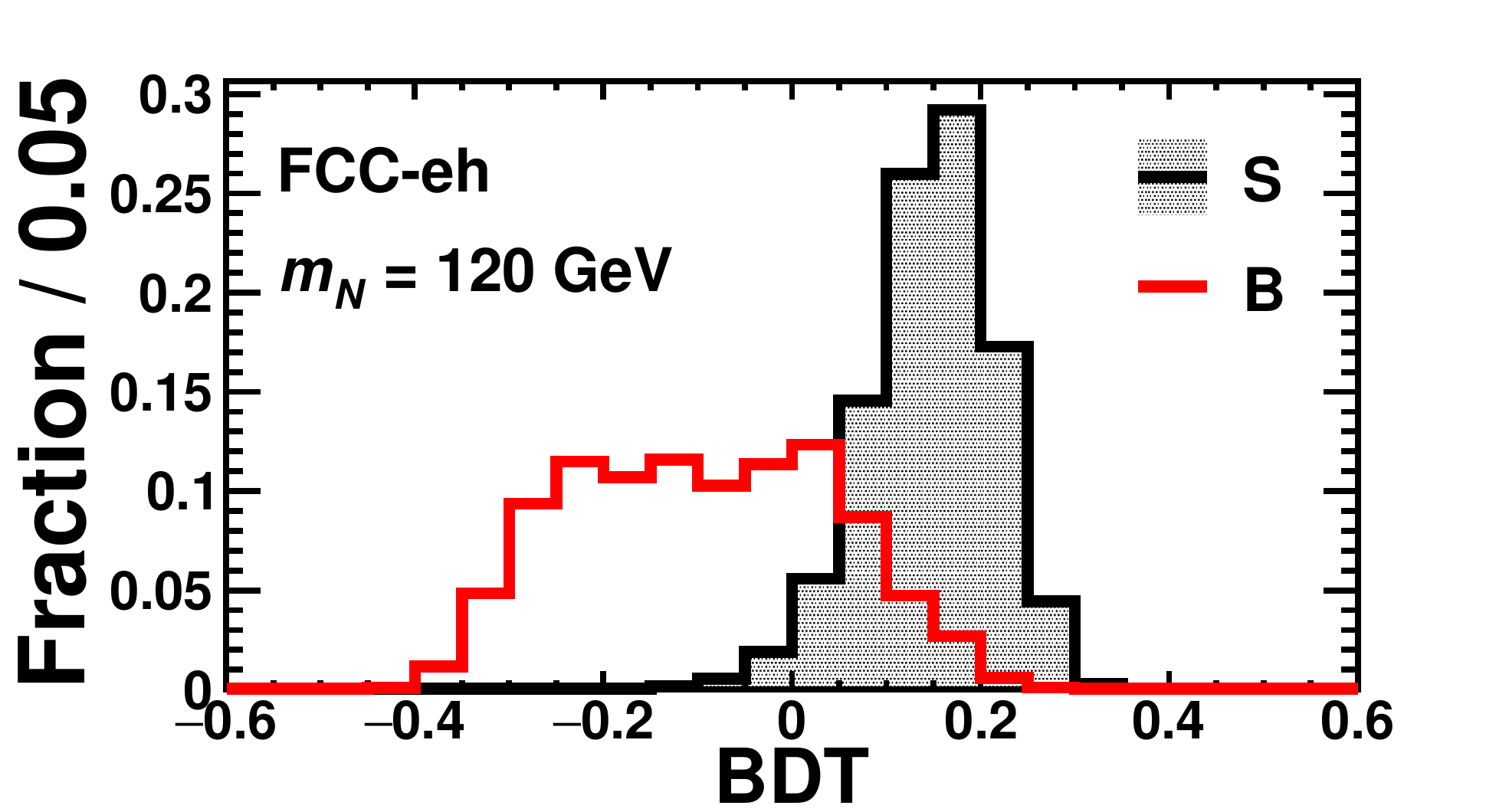}
\caption{
Distributions of BDT responses for the signal with $m_{N}$ = 120 GeV (black, filled) and total SM background (red) at the LHeC (left) and FCC-eh (right).
}
\label{fig:BDTbench}
\end{figure}
The Boosted Decision Tree (BDT) method in the TMVA package is adopted to separate the background from the signal.
Fig.~\ref{fig:BDTbench} shows the BDT distributions for the total background and the benchmark signal with $m_{N}$ = 120 GeV at the LHeC and FCC-eh. The BDT distributions of the signal and background are well separate, which means that a BDT cut can be applied to reject the background.
Comparing left and right plots, one sees that the signal distributions of LHeC and FCC-eh are similar, but the background distribution of LHeC has an obvious peak, while the background distribution of FCC-eh is flatter. In addition, the distributions of signal and background overlap slightly less at the FCC-eh, indicating that the separation between the signal and background is better than LHeC.

Since the kinematics of the signal varies with $m_N$, the distributions of BDT response also change with $m_N$. 
In Figs.~\ref{fig:BDTLHeC} and~\ref{fig:BDTFCC-eh}, we show the BDT distributions of four background processes and the signal corresponding to more representative heavy neutrino masses at the LHeC and FCC-eh.
We observe that as $m_N$ increases, the signal and background B1 ($\mu^+ \mu^- e^- j j j$) become more separate. However, the signal and the other three backgrounds overlap more and more as $m_N$ changes from 20 to 200 GeV, and then separate more and more as $m_N$ changes from 200 GeV to 1 TeV. 
When $m_N =$ 1 TeV, BDT distributions of all background processes tend to be similar and are almost completely separate from the signal. However, the limit for $m_N =$ 1 TeV is still restricted by its small signal cross section.
We note that 
when the kinematical distributions of background and signal are similar, it is difficult to distinguish between the signal and background,
leading to large overlap between their BDT distributions.
Therefore, the extent of separation between the BDT distributions of signal and background are determined by  the degree of deviation of their kinematics.

We note that the above input observables including the four-momenta and angles are very basic and usually called low-level variables for the MVA analysis. One can also construct some complicated observables and input such high-level variables to perform the MVA analysis.
To compare the effects on the final limits by inputting different sets of variables, we construct twenty-nine high-level observables and show the distributions of eight representative ones for the signal with $m_N = 120$ GeV and four background processes at the LHeC and FCC-eh in Fig.~\ref{fig:HLobsLHeC} and Fig.~\ref{fig:HLobsFCCeh}.
The representative observables are ordered according to the separation between the signal and background. At both colliders, the best observables to separate the signal from background are $p_{\rm T}(\mu)$, $E(\mu)$ and $\eta(\mu)$. This is mainly because the two main background processes B1 ($\mu^+ \mu^- e^- j j j$, blue) and B3 ($\mu^+ \nu_\mu e^- j j j$, pink) are well separate from the signal for these observables. One observes that the invariant mass $M(\mu+j_{2}+j_{3})$ can also be a good discriminator. This is because $M(\mu+j_{2}+j_{3})$ has a sharp peak around $m_N$, which means that it can also be used to reconstruct the heavy neutrino mass.
 
\begin{figure}[h]
\centering
\includegraphics[width=4cm,height=3cm]{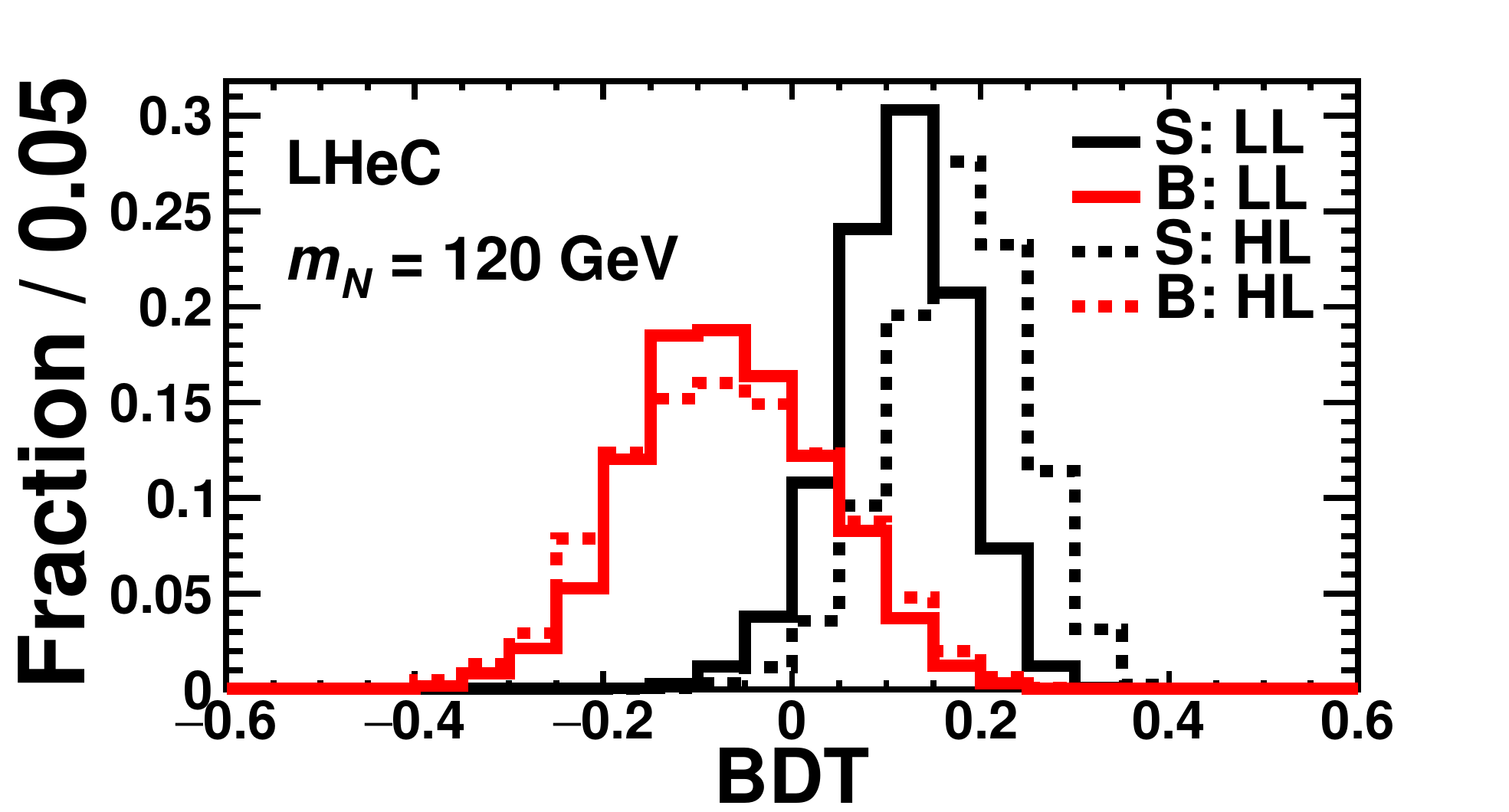}
\includegraphics[width=4cm,height=3cm]{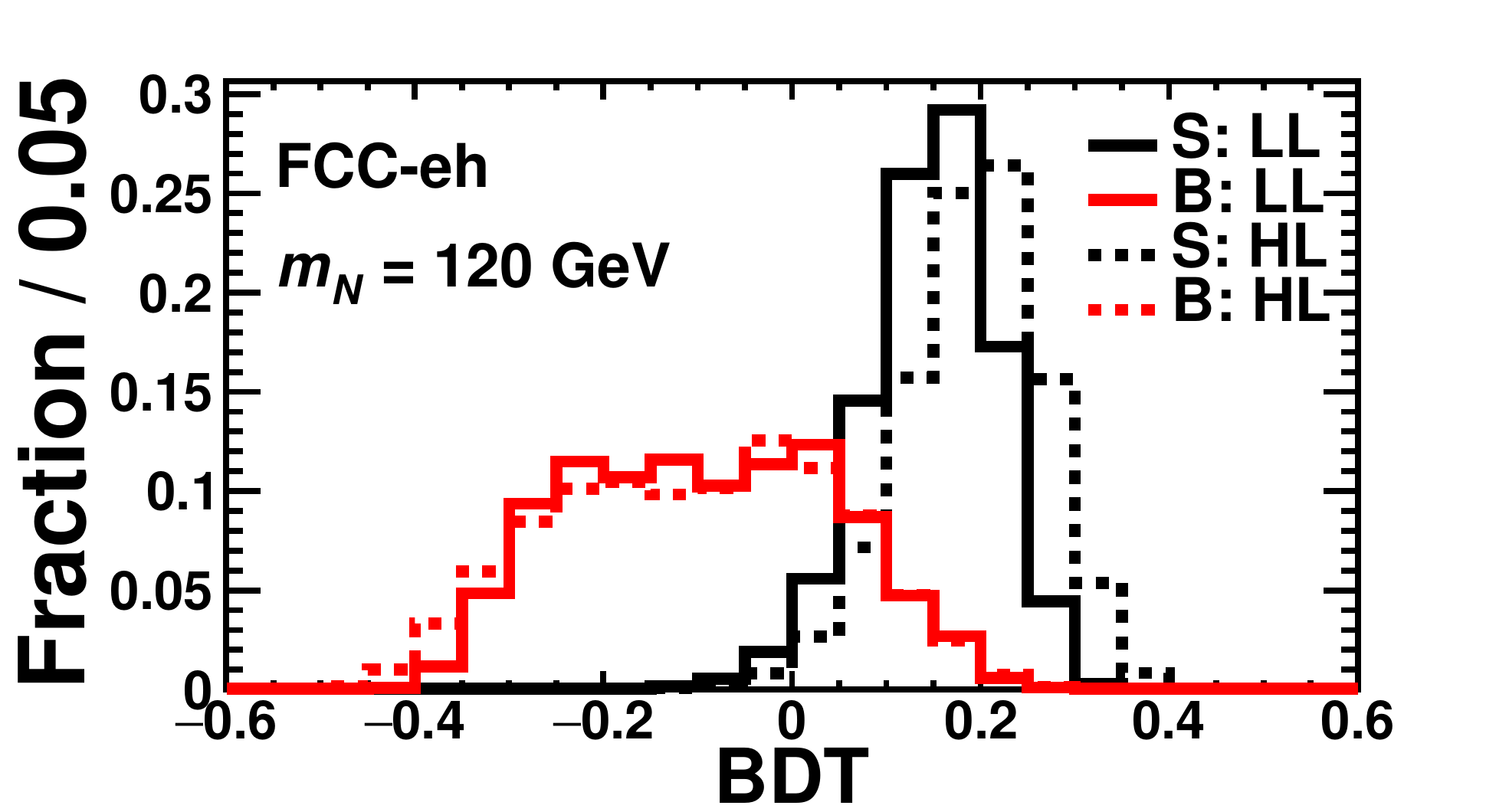}
\caption{
Distributions of BDT responses for the signal with $m_{N}$ =120 GeV and total background when inputting high-level (HL, dashed) and low-level (LL, solid) observables at the LHeC (left) and FCC-eh (right).
}
\label{fig:BDTcomparehl}
\end{figure}
Fig.~\ref{fig:BDTcomparehl} shows the distributions of BDT responses for the signal with $m_{N}$ =120 GeV and total background when inputting high-level (HL, dashed) and low-level  (LL, solid) observables at the LHeC (left) and FCC-eh (right).
One sees that the BDT distributions are similar between the low- and high-level cases.
Although some high-level observables seem to separate the signal better from the background than low-level observables, each high-level observable is not independent of each other and has correlations. 
The MVA-BDT analysis combines the information from all input observables with correlations. Since two sets of low- and high-level observables contain similar information, their BDT distributions should be similar as well.

To estimate the effects on final limits, we complete the analyses for the benchmark $m_N =$ 120 GeV using high-level observables at the LHeC (FCC-eh) and find that BDT cut efficiencies change from $7.71\,\,(7.95) \times 10^{-1}$ to $7.27\,\,(7.20) \times 10^{-1}$ for the signal and from $1.03 \times10^{-1}\,\, (8.85 \times10^{-2})$ to $3.95\,\, (3.17) \times10^{-2}$
for the total background. The final 2-$\sigma$ limit decreases slightly from $3.6\times10^{-6} $ to $ 2.4\times10^{-6} $ for LHeC and from $ 1.1\times10^{-6} $ to $ 7.3\times10^{-7} $ for FCC-eh, respectively.
Because the changes in the final results are small and more computing resources are needed to input more observables, for simplicity we input low-level variables to TMVA to obtain the results below.

\section{Results}
\label{sec:result}

In this section, based on our analyses we show the limits on the mixing parameter $|V_{\ell N }|^2$ for the heavy neutrino mass $m_N$ in the range of 10 to 1000 GeV.
After the pre-selection, the BDT cut is optimized according to the signal statistical significance calculated by Eq.~\ref{eqn:statSgf} for each mass case.
\begin{equation}
\sigma_{\rm stat} = \sqrt{2 [(N_s+N_b) {\rm ln}(1+\frac{N_s}{N_b}) - N_s ] },
\label{eqn:statSgf}
\end{equation}
where $N_s$ ($N_b$) is the number of signal (total background) events after all selection cuts.

In Table~\ref{tab:allEfficiencies}, we show selection efficiencies of pre-selection and BDT cuts for both signal and background processes at the LHeC and FCC-eh for representative heavy neutrino masses.
The total selection efficiency is the product of pre-selection and BDT cut efficiencies.
The number of background events after all cuts can be calculated by multiplying the initial number in Table~\ref{tab:cutmn120} by the total selection efficiency, while the number of signal events can be calculated as the product of signal cross section, collider luminosity and total selection efficiency.

\begin{figure}[h]
\centering
\includegraphics[width=8cm,height=6cm]{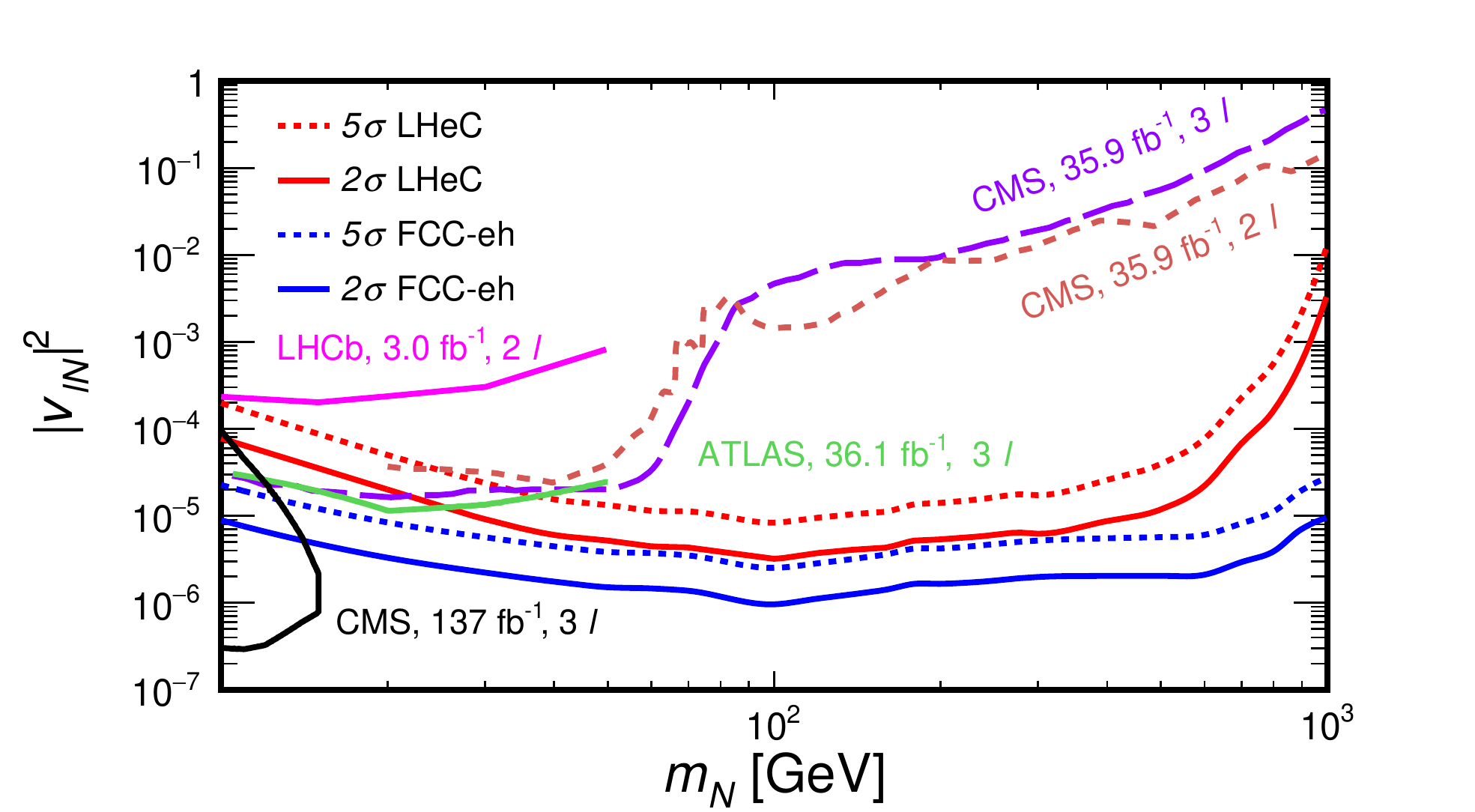}   
\caption{
2- and 5-$\sigma$ limits on mixing parameter $|V_{\ell N }|^2$ for the heavy neutrino mass in the range of 10 to 1000 GeV at the LHeC and FCC-eh. 
Also shown are the current experimental limits at 95\% confidence level from the trilepton searches (CMS, 35.9 $\ifb$, $3 \ell$)~\cite{CMS:2018iaf}, (CMS, 137 $\ifb$, $3 \ell$)~\cite{CMS:2021lzm},  (ATLAS, 36.1 $\ifb$, $3 \ell$)~\cite{ATLAS:2019kpx} and dilepton searches~(CMS, 35.9 $\ifb$, $2 \ell$)~\cite{CMS:2018jxx}, (LHCb, 3.0 $\ifb$, $2 \ell$)~\cite{LHCb:2020wxx} at the LHC.
}
\label{fig:sensitivity}
\end{figure}

In Fig.~\ref{fig:sensitivity}, we show 2- and 5-$\sigma$ limits on mixing parameter $|V_{\ell N }|^2$ for the heavy neutrino mass in the range of 10 to 1000 GeV at the LHeC (FCC-eh) with an electron beam energy of 60 GeV, a proton beam energy of 7 (50) TeV and an integrated luminosity of 1 (3) $\iab$.
At the LHeC, as $m_N$ changes from 10 GeV to 100 GeV, the 2-$\sigma$ upper limits on $|V_{\ell N }|^2$ decrease from $7.8 \times 10^{-5}$ to $3.2 \times 10^{-6}$; the limits are relatively flat for $m_N$ between 100 GeV to 500 GeV and increase rapidly to $3.3 \times 10^{-3}$ afterwards. 
The 2-$\sigma$ limit at the FCC-eh has similar behavior as that at the LHeC, but its varying range is much smaller, between $\sim 10^{-6}$ and $10^{-5}$.
At both colliders, the 5-$\sigma$ limits are slightly weaker than those for 2-$\sigma$.

The increasing or decreasing behavior of the upper limit is the result of the competition between the signal cross section and the separation extent of the signal and background kinematical distributions.  
With the increase of $m_N$, the signal cross section decreases gradually as shown in Fig.~\ref{fig:crs}, while BDT distributions of the signal and the total background become more and more separate (cf. Figs.~\ref{fig:BDTLHeC} and~\ref{fig:BDTFCC-eh}).
Therefore, limits are weaker because of smaller separations between the signal and background kinematics for small masses and because of smaller signal cross sections for heavy masses.
The most stringent limits are achieved at $m_{N} \sim 100$ GeV.
Limits are relatively flat in the middle range of $m_N$ because of the offset between the decrease of the signal cross section and increase of the separation extent of the signal and background kinematics.
At the LHeC, when $m_{N}$ $\gtrsim$ 500 GeV, the signal cross section decreases rapidly with the increase of $m_{N}$, leading to the rapid increase in the upper limit.
Since the signal cross section at the FCC-eh decreases slowly, the limit for heavy masses also increases gently.
The FCC-eh has better limits than the LHeC, mainly because the signal cross section of FCC-eh is larger than that of LHeC for the same $m_{N}$.

To compare with current experiment limits, we also present the recent LHC limits in Fig.~\ref{fig:sensitivity}.
The details of these studies are reviewed in Sec.~\ref{sec:intro}.
The limit curve for (CMS, 35.9 $\ifb$, $2 \ell$) is reinterpreted from the original limit on the parameter $|V_{eN }V_{\mu N }^*|^2/(|V_{eN }|^2+|V_{\mu N }|^2)$ in the CMS same-sign dilepton search~\cite{CMS:2018jxx}. With the same assumption of $|V_{eN }|^2 = |V_{\mu N }|^2$ in this article, the original limit is shown by multiplying a factor of two.
The limit for (CMS, 137 $\ifb$, $3 \ell$) is from the Ref.~\cite{CMS:2021lzm} where the CMS collaboration searched for a long-lived heavy neutrino in the trilepton final state with displaced vertices.
It has excluded some parameter regions when $m_N < 15$ GeV.
Our analyses show that the LHeC limit is slightly weaker than the current CMS and ATLAS trilepton searches when $m_N$ $\lesssim$ 20 GeV, while it is better for heavier masses and about two to three orders of magnitude stronger than the current CMS limits when $m_N$ $\gtrsim$ 100 GeV.
Compared with current LHC limits, the FCC-eh gives more stringent limits when $m_N > 15$ GeV and are much stronger when $m_N$ $\gtrsim$ 100 GeV.

\section{Summary and Discussion}
\label{sec:sum}

In this paper, we utilize the lepton number violation signal process of $p\, e^- \to \mu^+ jjj$ to search for heavy Majorana neutrinos at the future electron-proton colliders.
We consider the LHeC (FCC-eh) running with an electron beam energy of 60 GeV, a proton beam energy of 7 (50) TeV and an integrated luminosity of 1 (3) $\iab$.
To simplify the analyses, we consider the simplified Type-I model and assume that only one generation of heavy neutrinos $N$ is within the collider access and mixes with active neutrinos of electron and muon flavours with the same mixing parameters, i.e. $|V_{\ell N}|^2 = |V_{e N}|^2 = |V_{\mu N}|^2$ and $|V_{\tau N}|^2 = 0$.
The signal production cross sections are presented in Fig.~\ref{fig:crs} at both LHeC and FCC-eh for the heavy neutrino mass $m_N$ in the range of 10$-$1000 GeV.
We apply detector configurations and simulate signal and four dominant SM background events of $\mu^+ \mu^- e^- j j j$, $\mu^+ \mu^- \nu_e j j j$, $\mu^+ \nu_\mu e^- j j j$ and $\mu^+ \nu_\mu \nu_e j j j$ including detector effects. 

We first use the pre-selection cuts to select the final state with exactly one muon with positive charge, at least three jets and small missing energy. 
The number of events for the signal with benchmark $m_N =$ 120 GeV and four background processes after applying pre-selection cuts are presented in Table~\ref{tab:cutmn120}.
To reject the background efficiently, nineteen basic observables are input to perform the multi-variate analyses based on machine-learning. The distributions of BDT responses of the signal and the SM background processes at the LHeC and FCC-eh for representative heavy neutrino masses are shown in Appendix~\ref{appendix:BDTdistribution}, while the efficiencies of pre-selection and BDT cuts are shown in Appendix~\ref{appendix:efficiencies}.

To test the effects on the final limits by inputting different sets of observables, we construct and input another set of twenty-nine observables and find that the changes in the final limits are small. The distributions of eight high-level representative kinematic observables are also presented in Appendix~\ref{appendix:HLobs} for the signal with $m_N =$ 120 GeV and four background processes at the LHeC and FCC-eh.

Based on our analyses, we show the 2- and 5-$\sigma$ upper limits on the mixing parameter $|V_{\ell N}|^2$ for the heavy neutrino mass $m_N$ in the range of 10$-$1000 GeV at both LHeC and FCC-eh in Fig.~\ref{fig:sensitivity}.
At the LHeC, the 2-$\sigma$ upper limits on $|V_{\ell N }|^2$ decrease from $7.8 \times 10^{-5}$ to $3.2 \times 10^{-6}$ when $m_N$ changes from 10 GeV to 100 GeV; the limits are relatively flat for $m_N$ in the middle range between 100 GeV and 500 GeV and increase rapidly to $3.3 \times 10^{-3}$ afterwards. 
The 2-$\sigma$ limit at the FCC-eh has the similar behavior as that at the LHeC, but its varying range is much smaller, between $\sim 10^{-6}$ and $10^{-5}$.
At both colliders, the 5-$\sigma$ limits are slightly weaker than those for 2-$\sigma$.

The limits are compared with the current LHC experimental limits. Our analyses show that the LHeC limit is slightly weaker than the current CMS and ATLAS trilepton searches when $m_N$ $\lesssim$ 20 GeV, while it is better for heavier masses and about two to three orders of magnitude stronger than the current CMS limits when $m_N$ $\gtrsim$ 100 GeV.
Compared with current LHC limits, the FCC-eh gives more stringent limits when $m_N > 15$ GeV and are much stronger when $m_N$ $\gtrsim$ 100 GeV.

The LNV signal process considered in this study is one typical channel to search for heavy Majorana neutrinos at electron-proton colliders. 
If this signal is discovered at colliders, it is also confirmed that the nature of $N$ is of Majorana type. 
Therefore, our results are an important complement to the physics goals of electron-proton colliders, and the parameter space probed by our search strategy could explain other fundamental physics problems, such as the leptogenesis~\cite{Drewes:2021nqr}.

We assume that $N$ decays promptly in this study. However,
when $N$'s lifetime is long enough, heavy neutrinos can have sizeable probability to travel through and decay outside the detector. 
In this case, such neutrinos will be mis-detected and behave as missing energy, and thus these events cannot  contribute to our signal.
The number of detectable signal events needs to be multiplied by the average probability $\bar{P}$ of heavy neutrinos decaying inside the detector's fiducial volume. 
Because our signal final state has one $\mu^+$ and two jets from the the decay of $N$, in order to detect these jets, the heavy neutrino needs to decay before the end of the hadronic calorimeter (HCAL). 

To estimate this effect on final limits, we consider detector layouts of the LHeC and FCC-eh  from Ref.~\cite{AbelleiraFernandez:2012ni} and Ref.~\cite{FCCeh:detector}, respectively.
We generate the signal sample $p e^- \to j N$ at the parton level using MadGraph for different $m_N$ at the LHeC and FCC-eh, and calculate $\bar{P}$ of detectors using a similar method used in Ref.~\cite{Wang:2019orr}.
Results for $\bar{P}$ while varying  $|V_{lN}|^2$ when $m_N = 5, 10, 20$ GeV at the LHeC and FCC-eh are derived, and we find that the resulting curves of detectors at LHeC and FCC-eh are very similar.
For $m_N = 5, 10, 20$ GeV, the $\bar{P}$ values begin to reduce from unity when the mixing parameter $|V_{\ell N}|^2 \sim 2 \times 10^{-6}, 3 \times 10^{-8}, 8 \times 10^{-10}$, respectively, and become zero when $|V_{\ell N}|^2 \sim 2 \times 10^{-9}, 5 \times 10^{-11}, 1 \times 10^{-12}$, respectively.
Since Fig.~\ref{fig:sensitivity} shows that the 2-$\sigma$ upper limits on $|V_{\ell N}|^2$ for $m_N = $ 10 GeV at the LHeC and FCC-eh are $7.8 \times 10^{-5}$ and $9.0 \times 10^{-6}$, respectively, which are larger than $3 \times 10^{-8}$, the corresponding $\bar{P}$ values are unity, and thus the final limits are not modified. 
For larger $m_N$, $|V_{\ell N}|^2$ is required to be even smaller such that $\bar{P}$ can decrease from unity. 
Therefore, the effect of long-lived cases of heavy neutrinos is expected to be negligible for the mass range considered in this study.
Such effect becomes significant for very small masses, and we leave it for future studies.

We note that when $|V_{e N}|^2 = |V_{\mu N}|^2$ the LNV signal $p\, e^- \to e^+ jjj$ also exists with  cross section approximately equal to that of signal $p\, e^- \to \mu^+ jjj$. 
However, for the signal $p\, e^- \to e^+ jjj$, there exists one additional SM background process of $p\, e^- \to e^- jjj$. When the final state $e^-$ is mis-detected as $e^+$, this process can be one significant background.
To estimate sensitivities on $|V_{\ell N}|^2$ of the $e^+$ final state signal, 
we consider its corresponding SM background processes.
The cross sections of background processes of $e^- jjj$, $e^+ e^- e^- j j j$, $e^+ e^- \nu_e j j j$ and $e^+ \nu_e \nu_e  j j j$ are found to be $1.4 \times 10^{4}$, $0.53$, $0.33$ and $8.1 \times 10^{-6}$ pb at the LHeC, respectively. 
One sees that the production cross section of $e^- jjj$ is much larger than the other background processes. Therefore, it could still dominate even if the detector's charge misidentification rate is small.

With the current detector configuration at the LHeC, after selecting exactly one positron with transverse momentum above 5 GeV, the total background cross section for the $e^+$ final state is $8.62\times10^{-2}$ pb, among which 
the $e^- jjj$ process is $6.07 \times 10^{-2}$ pb and the sum of the other three background processes is $2.55 \times 10^{-2}$ pb, so $e^- jjj$ is the main background.
For the $\mu^+$ final state, the total background cross section after selecting exactly one muon with positive charge with transverse momentum above 5 GeV is $2.58 \times 10^{-2}$ pb (cf. cut(i) in Table~\ref{tab:cutmn120}).

Because the total background cross section for the $e^+$ final state is about three times larger than that for the $\mu^+$ final state, the limit on mixing parameter $|V_{\ell N}|^2$ from the signal process $p\, e^- \to e^+ jjj$ is expected to be weaker than that from the signal process $p\, e^- \to \mu^+ jjj$. 
However, when $|V_{e N}|^2 \neq 0$ and $|V_{\mu N}|^2 = 0$, the $p\, e^- \to \mu^+ jjj$ signal cannot be produced any more, but the $p\, e^- \to e^+ jjj$ signal can still exist. Because the $p\, e^- \to e^+ jjj$ signal process depends on the mixing parameter $|V_{e N}|^2$ only, it can be a unique channel to probe $|V_{e N}|^2$ independent of other mixing parameters.
In this sense, the detailed analyses of the $e^+$ final state are still meaningful and we leave it for future studies.

\appendix

\section{Distributions of BDT responses}
\label{appendix:BDTdistribution}

In Fig.~\ref{fig:BDTLHeC} and Fig.~\ref{fig:BDTFCC-eh} , we show the distributions of BDT responses of the signal and SM background processes at the LHeC and FCC-eh with different $m_N$ assumptions. 

\begin{figure}[H]
\centering
\subfigure{
\includegraphics[width=4cm,height=3cm]{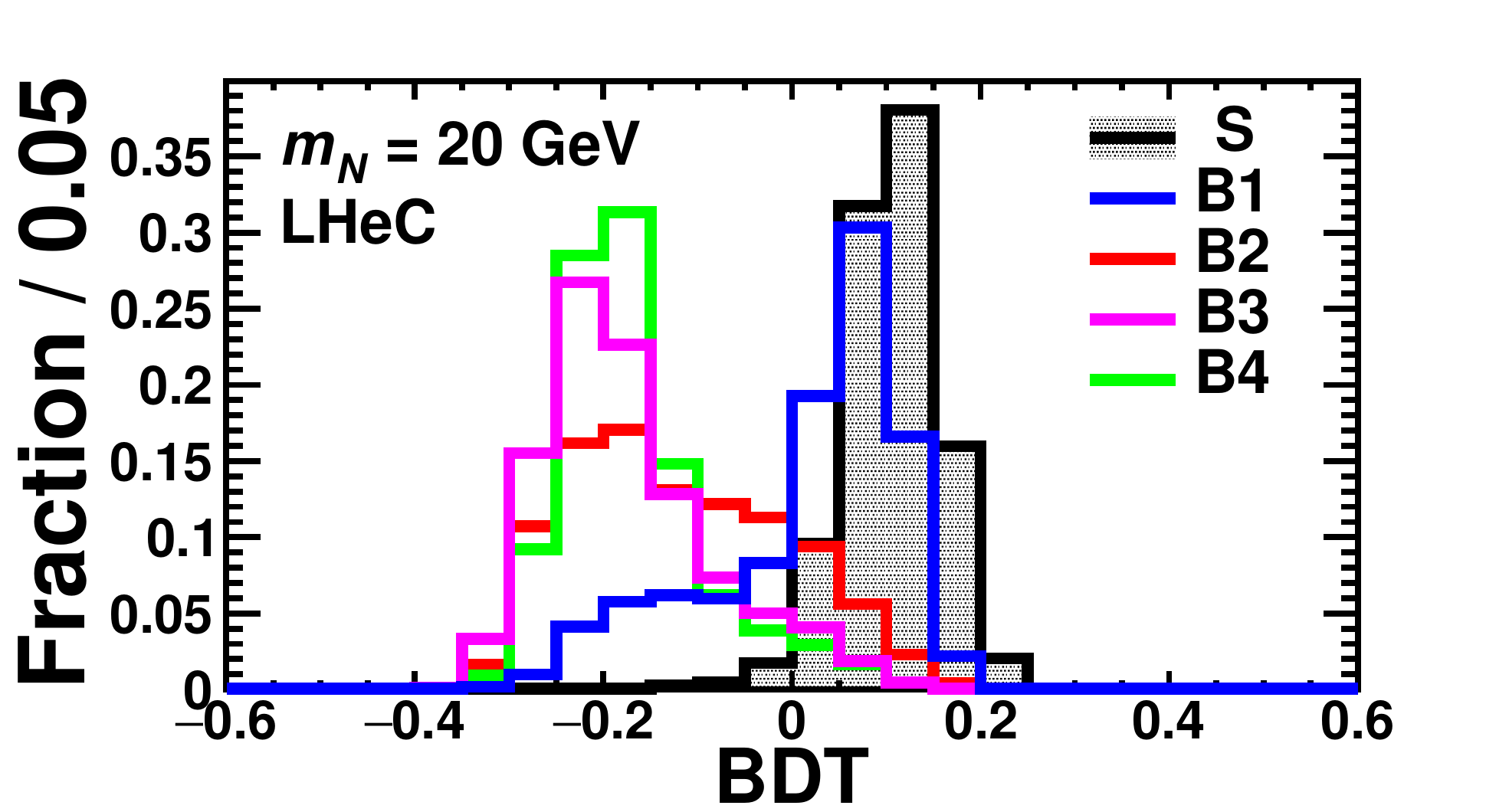}
\includegraphics[width=4cm,height=3cm]{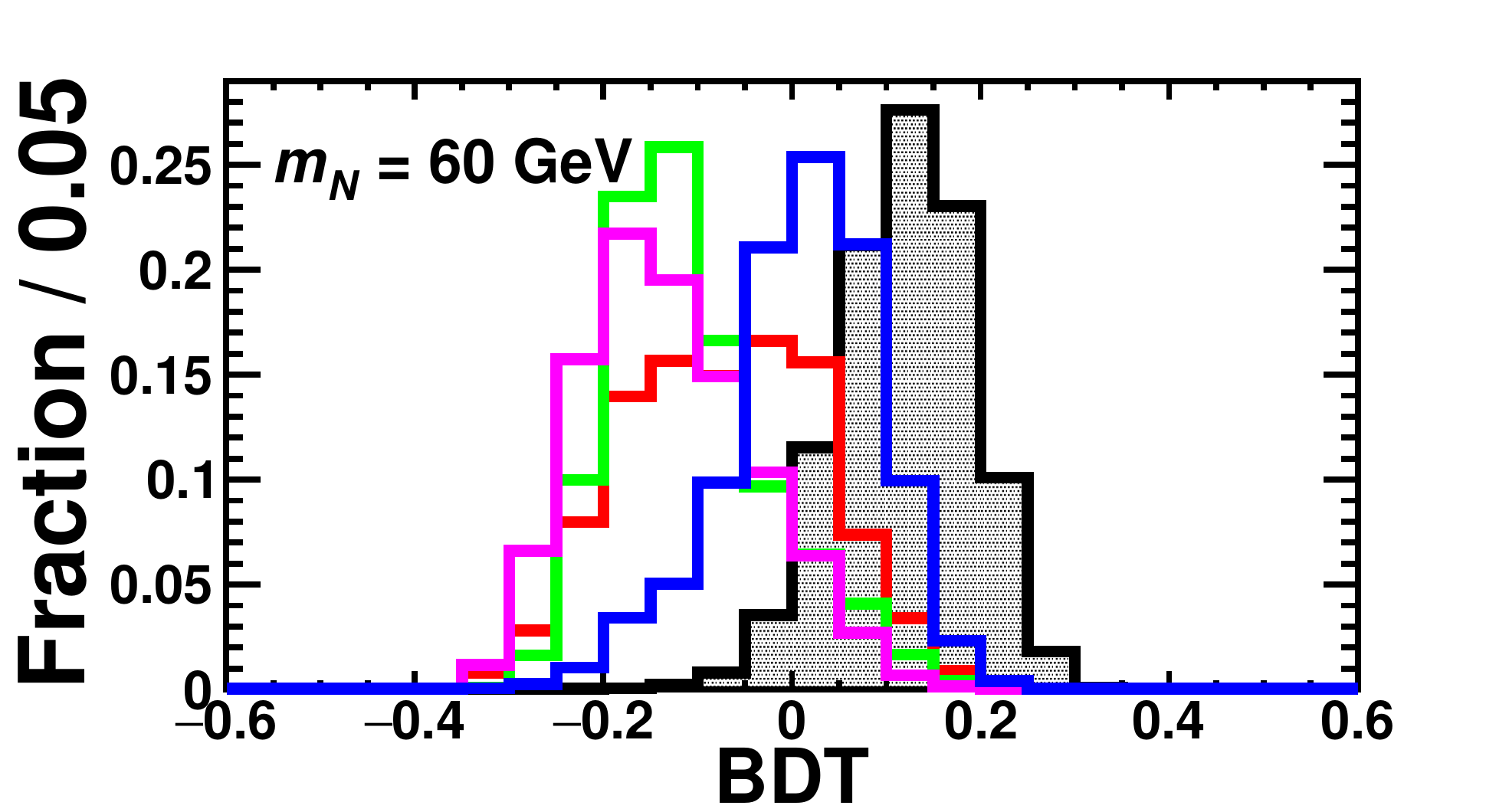}
}
\end{figure}
\addtocounter{figure}{-1}
\vspace{-1.0cm}
\begin{figure}[H]
\centering
\addtocounter{figure}{1}
\subfigure{
\includegraphics[width=4cm,height=3cm]{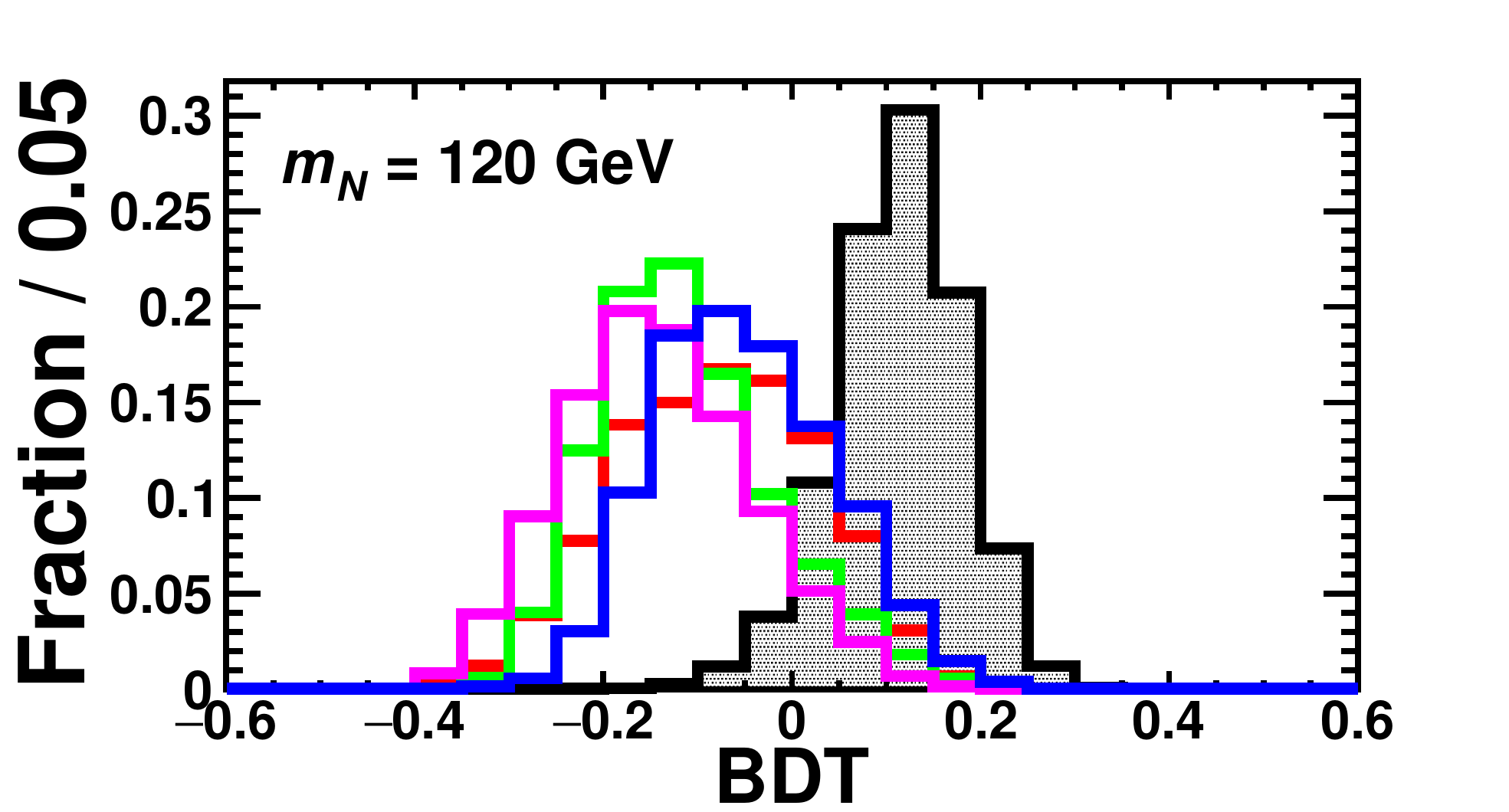}
\includegraphics[width=4cm,height=3cm]{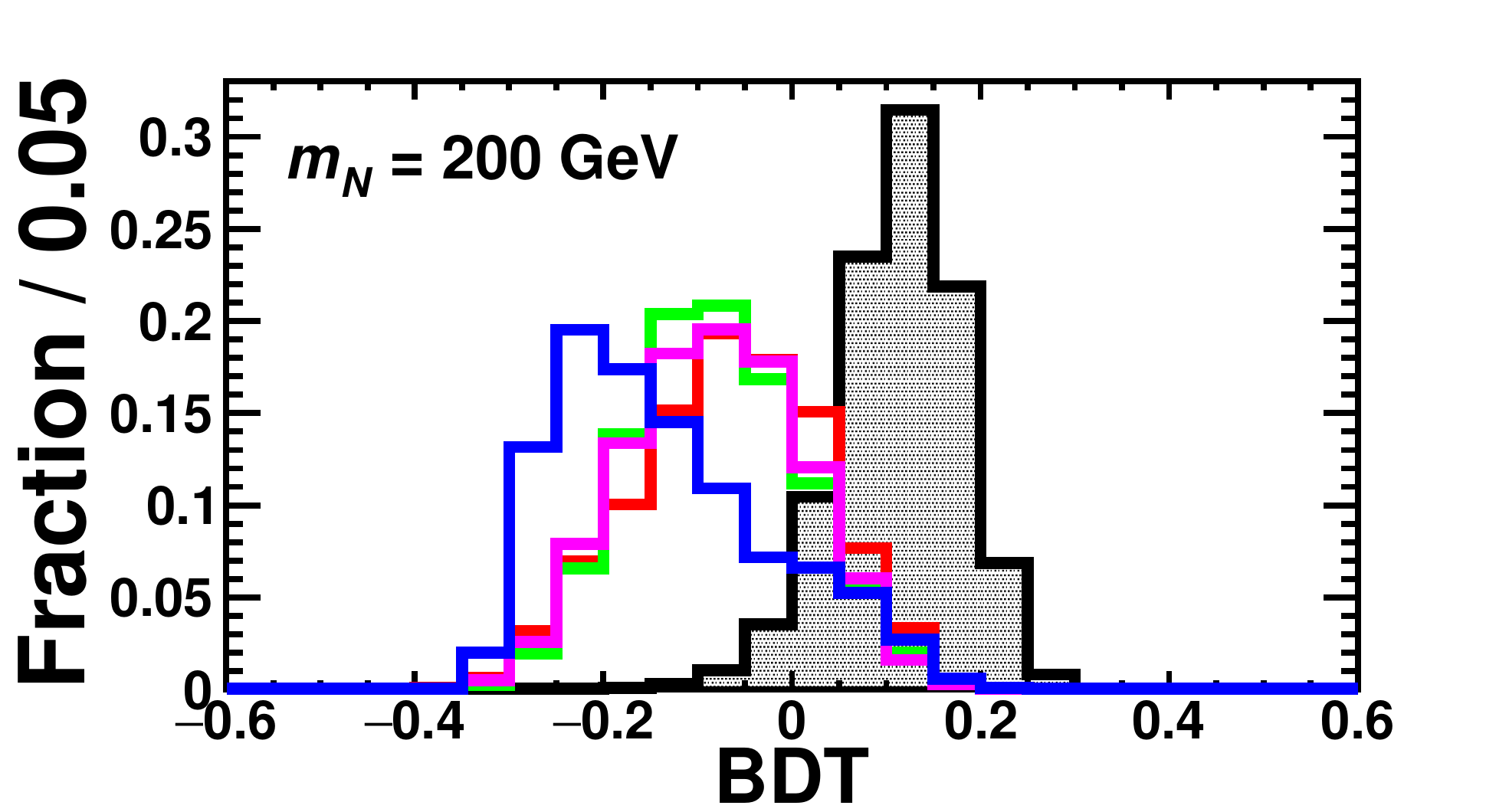}
}
\end{figure}
\vspace{-1.0cm}
\begin{figure}[H] 
\centering
\addtocounter{figure}{-1}
\subfigure{
\includegraphics[width=4cm,height=3cm]{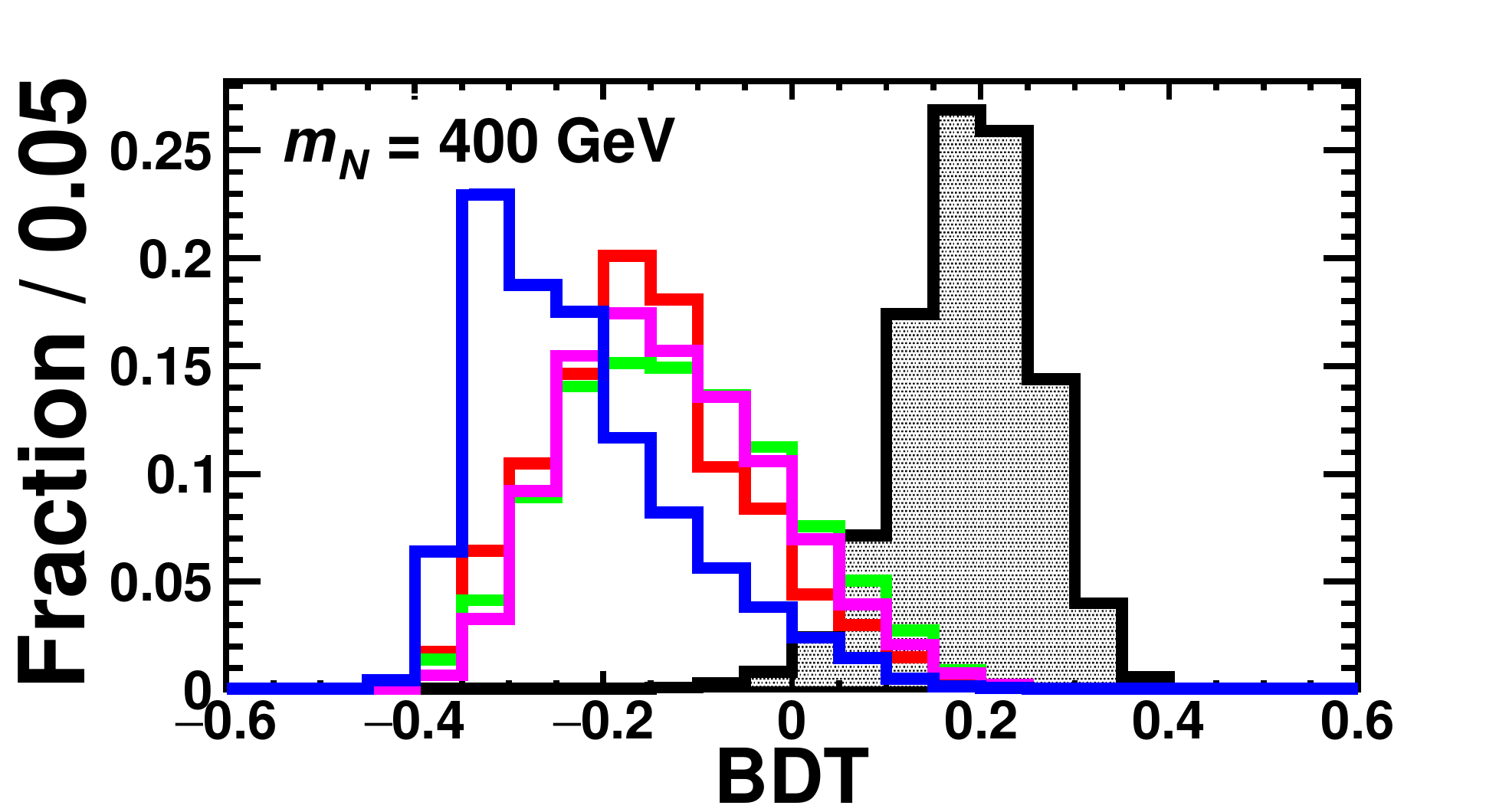}
\includegraphics[width=4cm,height=3cm]{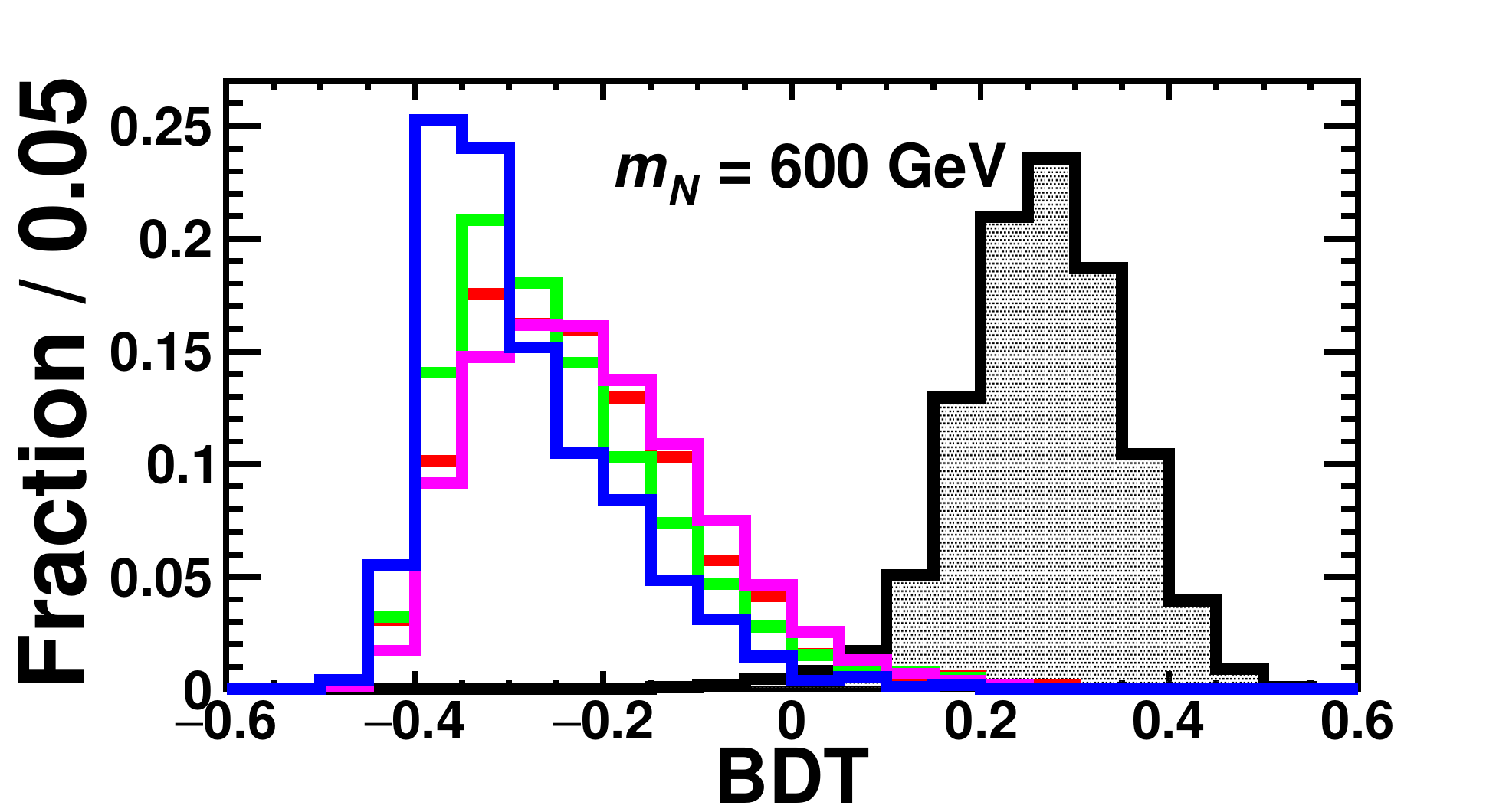}
}
\end{figure}
\vspace{-1.0cm}
\begin{figure}[H] 
\centering
\addtocounter{figure}{1}
\subfigure{
\includegraphics[width=4cm,height=3cm]{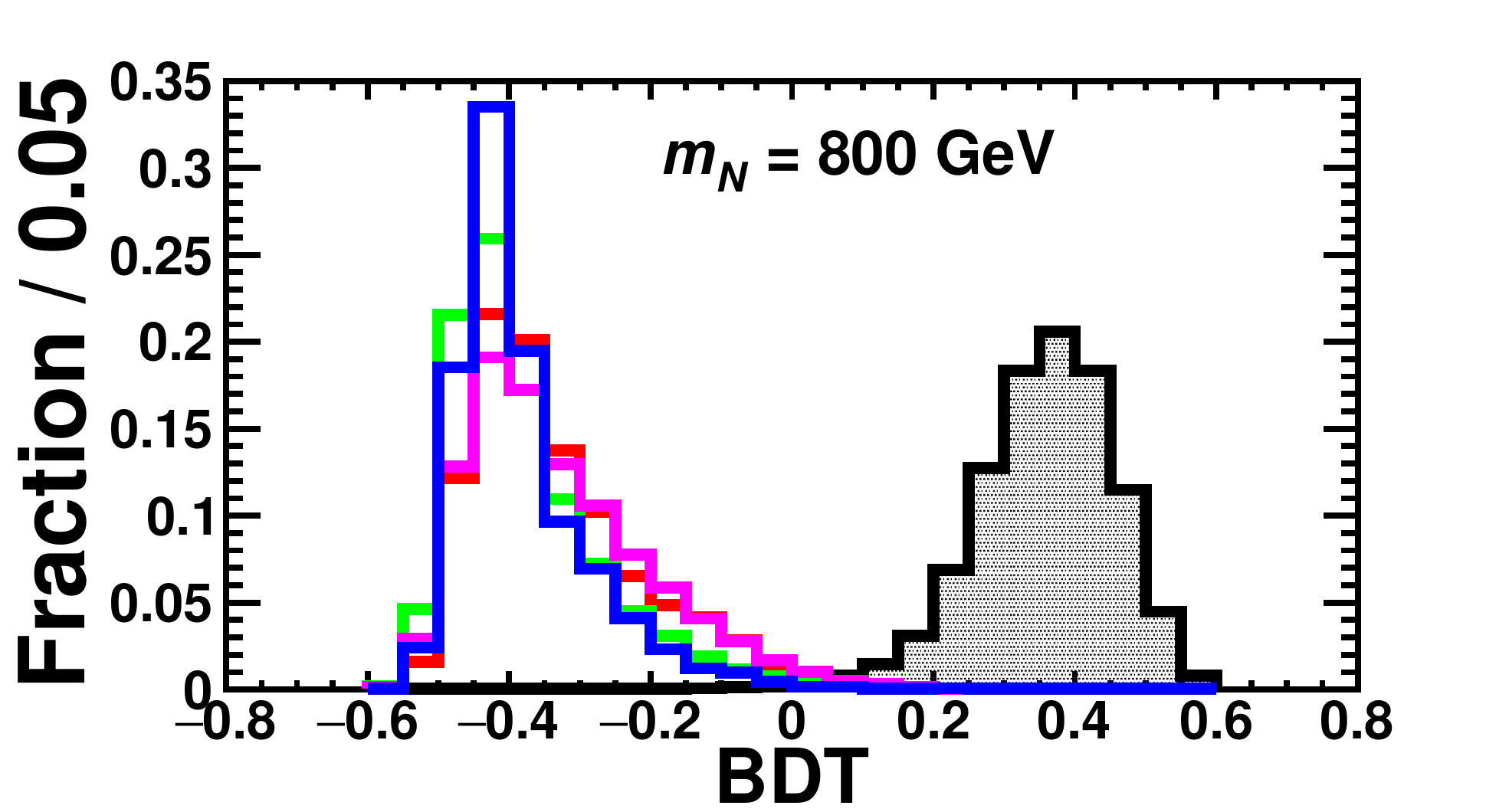}
\includegraphics[width=4cm,height=3cm]{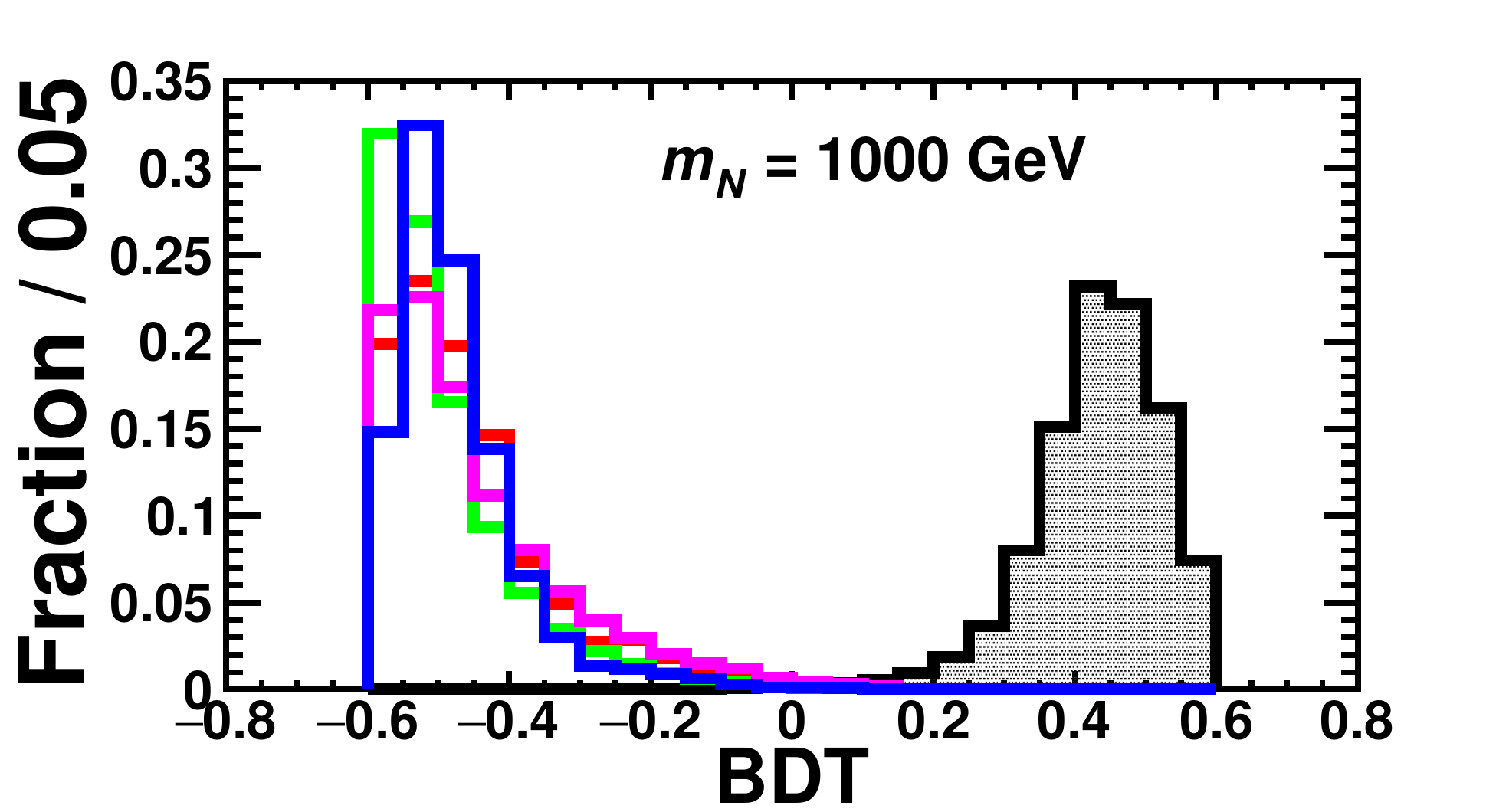}
}
\caption{Distributions of BDT responses of the signal (black, filled) and four background processes at the LHeC for representative heavy neutrino masses.}
\label{fig:BDTLHeC}
\end{figure}

\begin{figure}[H] 
\centering
\subfigure{
\includegraphics[width=4cm,height=3cm]{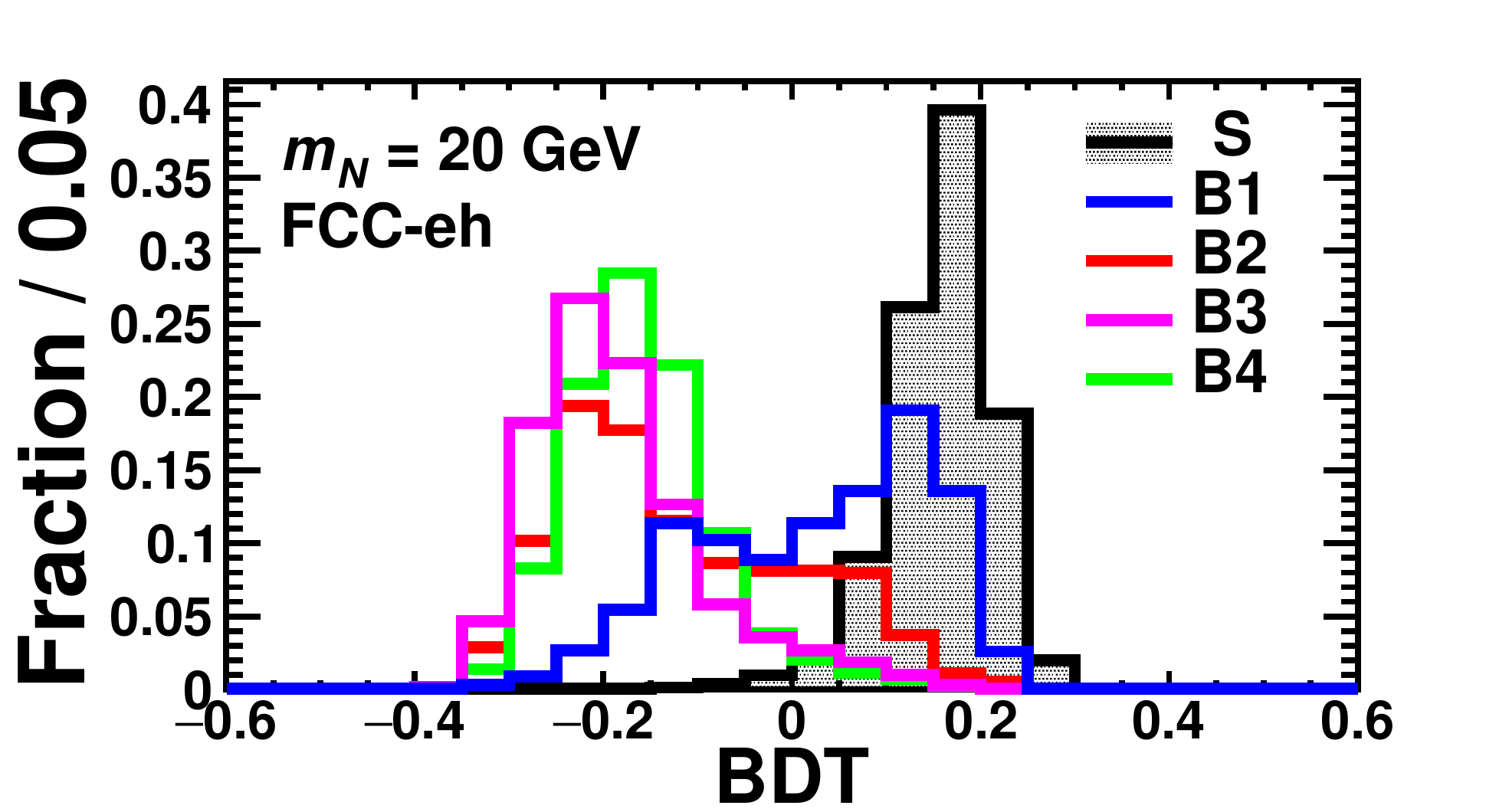}
\includegraphics[width=4cm,height=3cm]{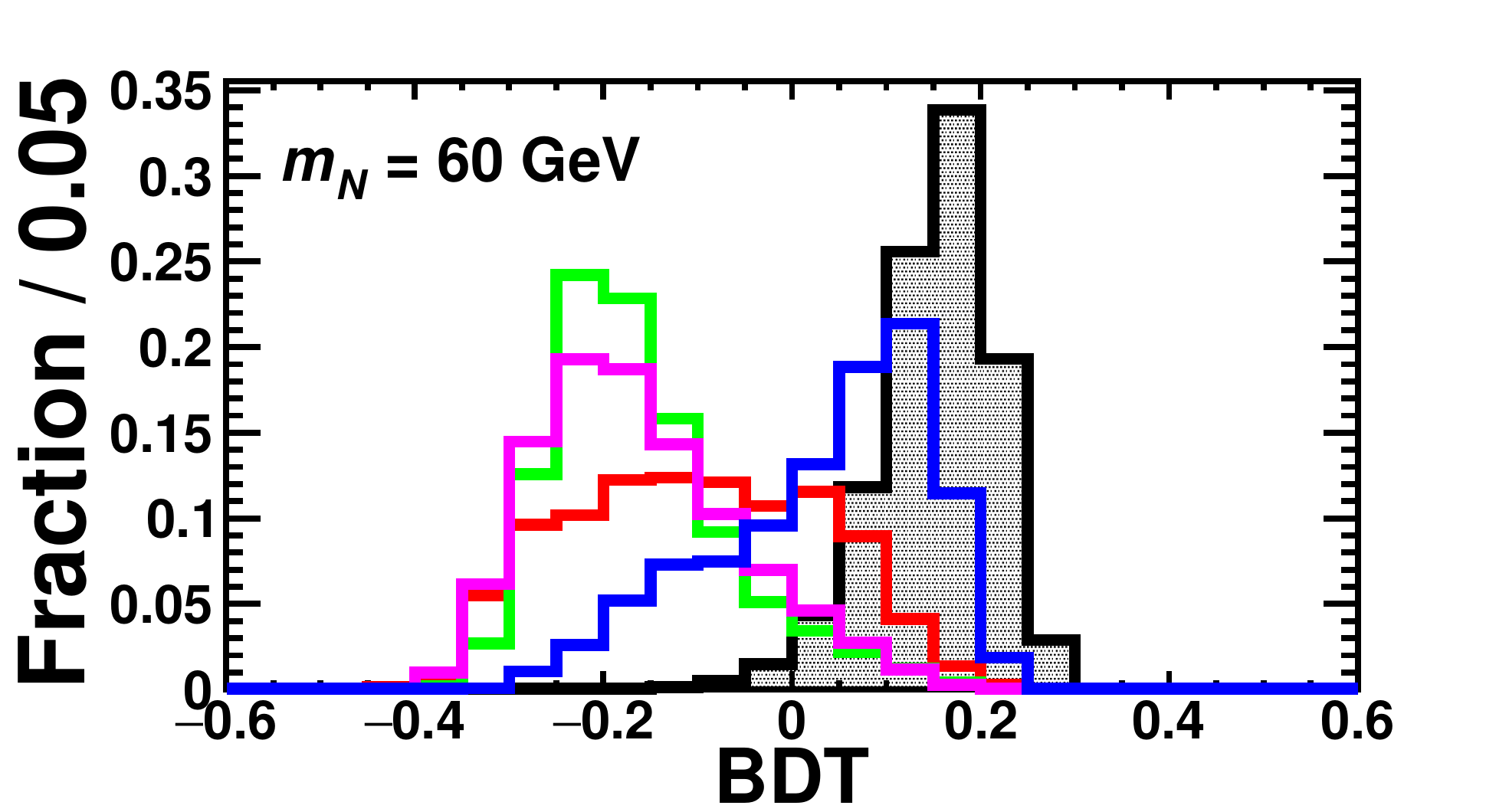}
}
\end{figure}
\addtocounter{figure}{-1}
\vspace{-1.0cm}
\begin{figure}[H] 
\centering
\addtocounter{figure}{1}
\subfigure{
\includegraphics[width=4cm,height=3cm]{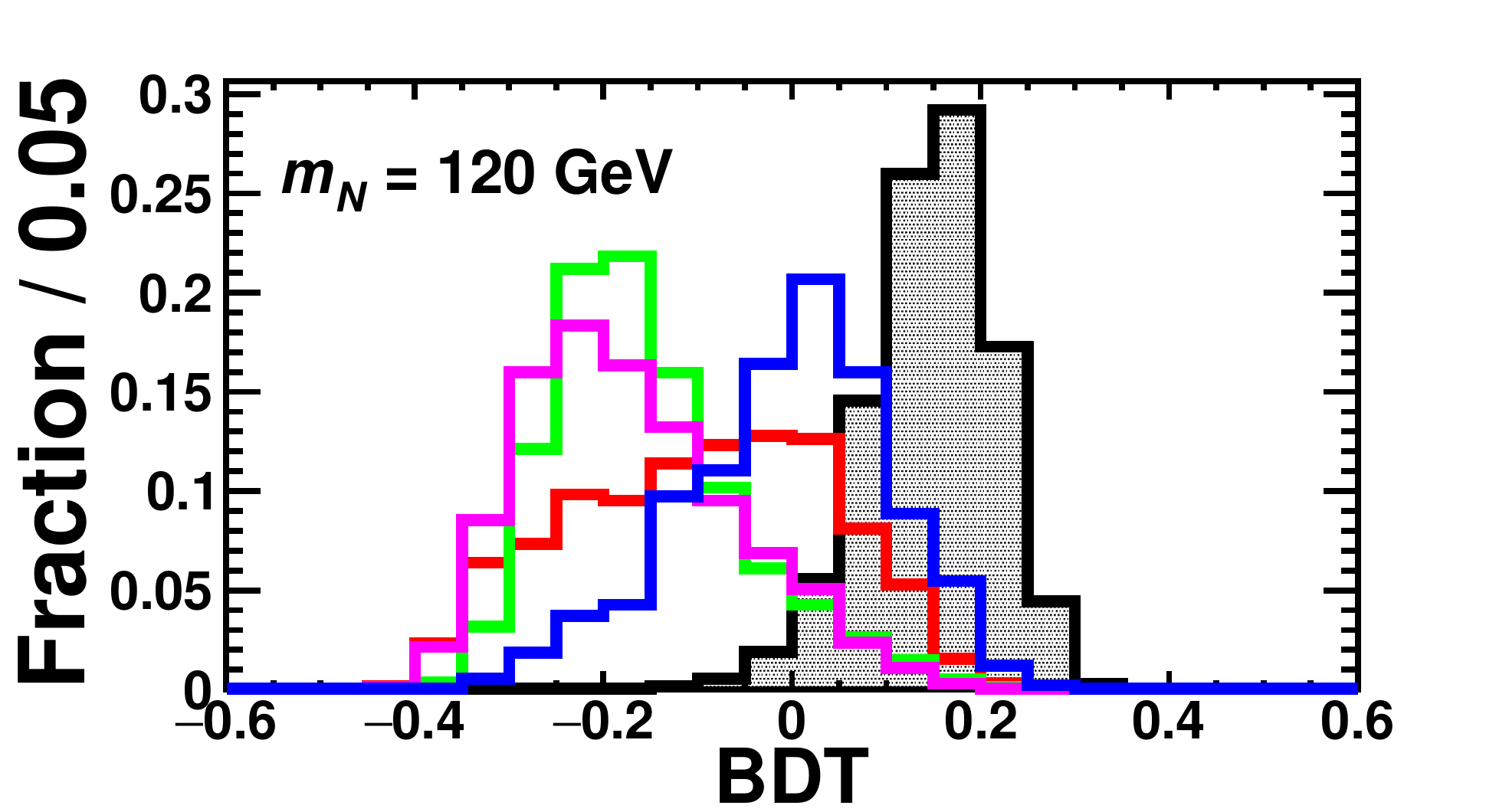}
\includegraphics[width=4cm,height=3cm]{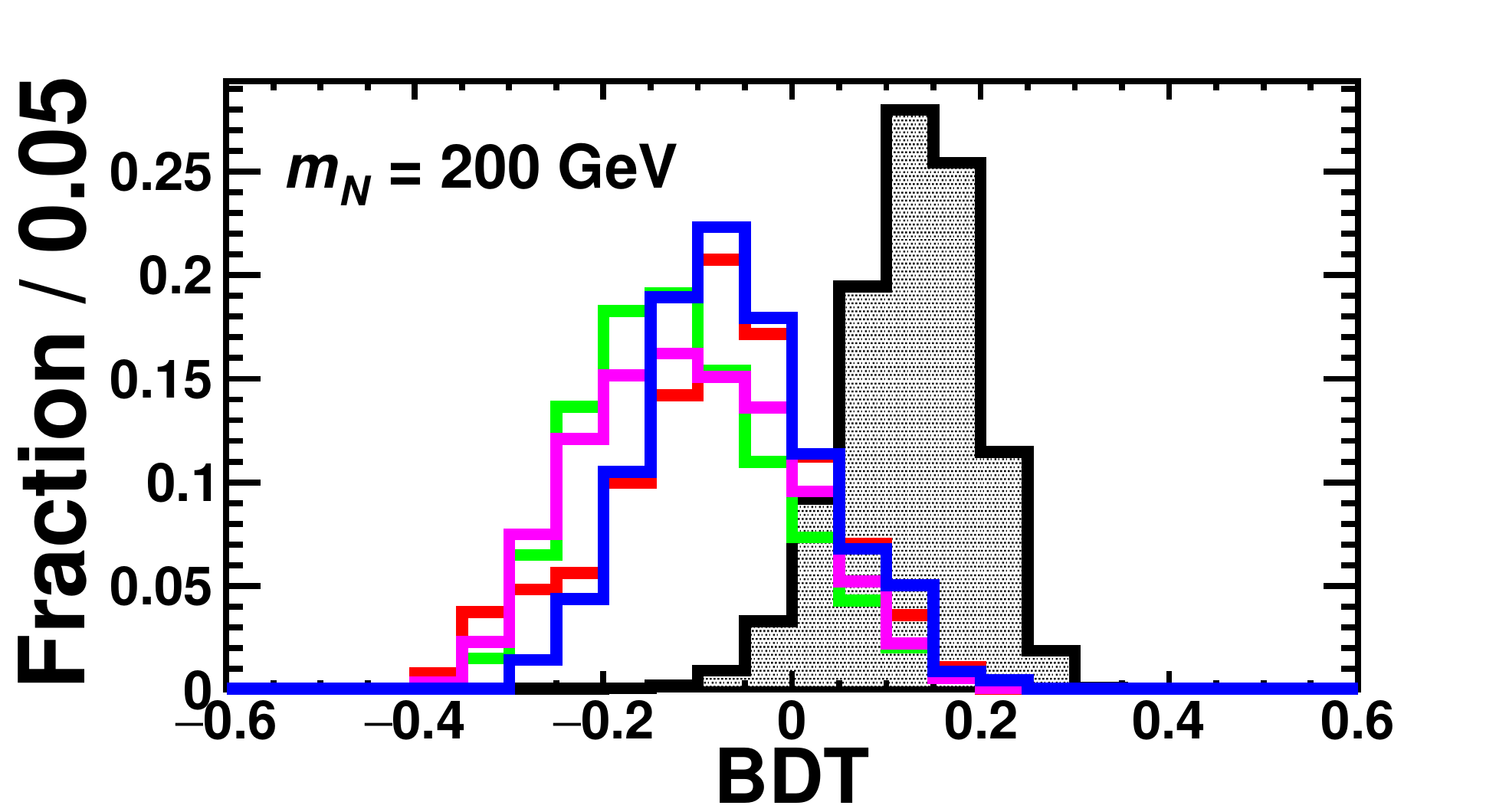}
}
\end{figure}
\vspace{-1.0cm}
\begin{figure}[H] 
\centering
\addtocounter{figure}{-1}
\subfigure{
\includegraphics[width=4cm,height=3cm]{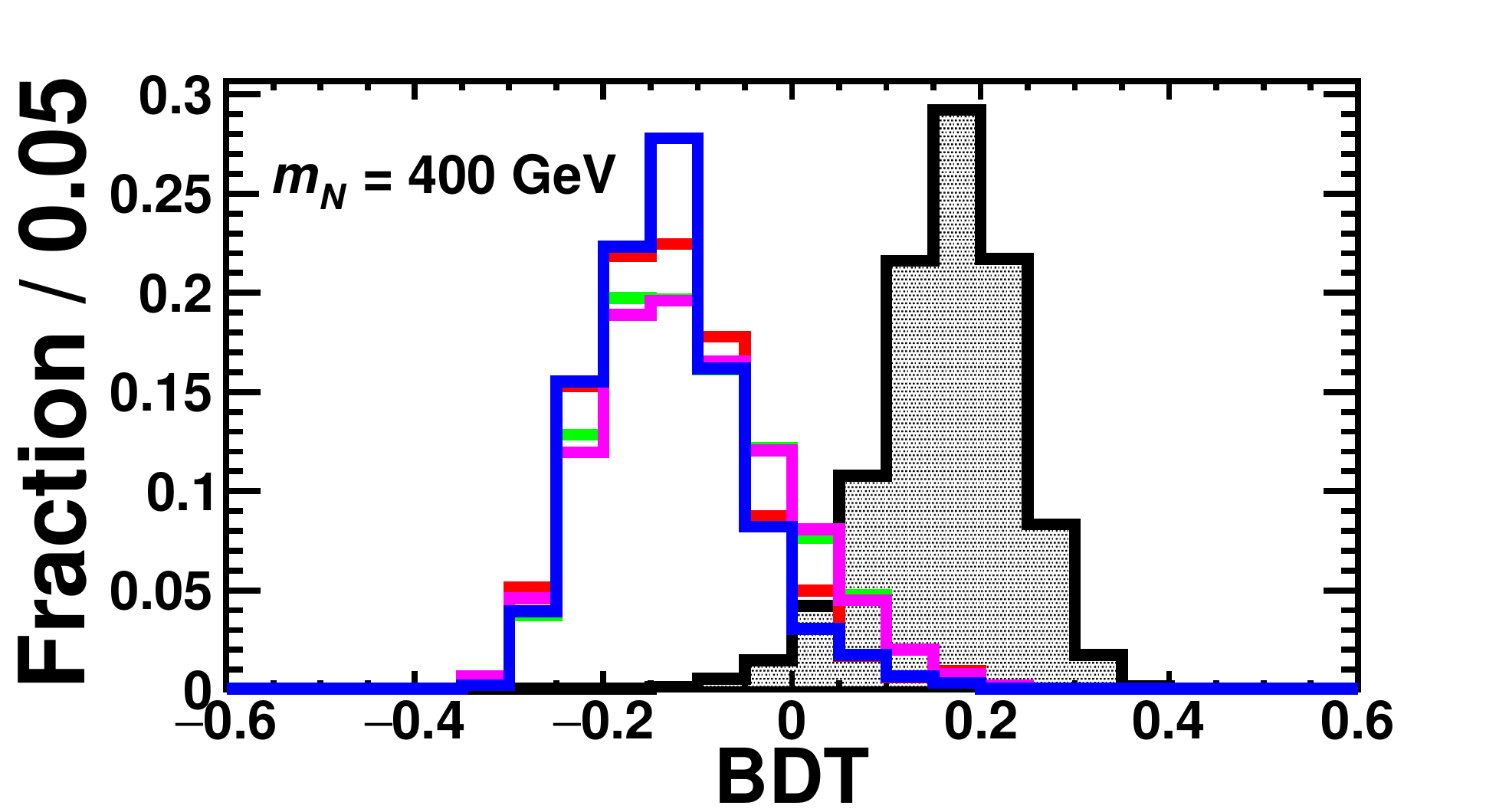}
\includegraphics[width=4cm,height=3cm]{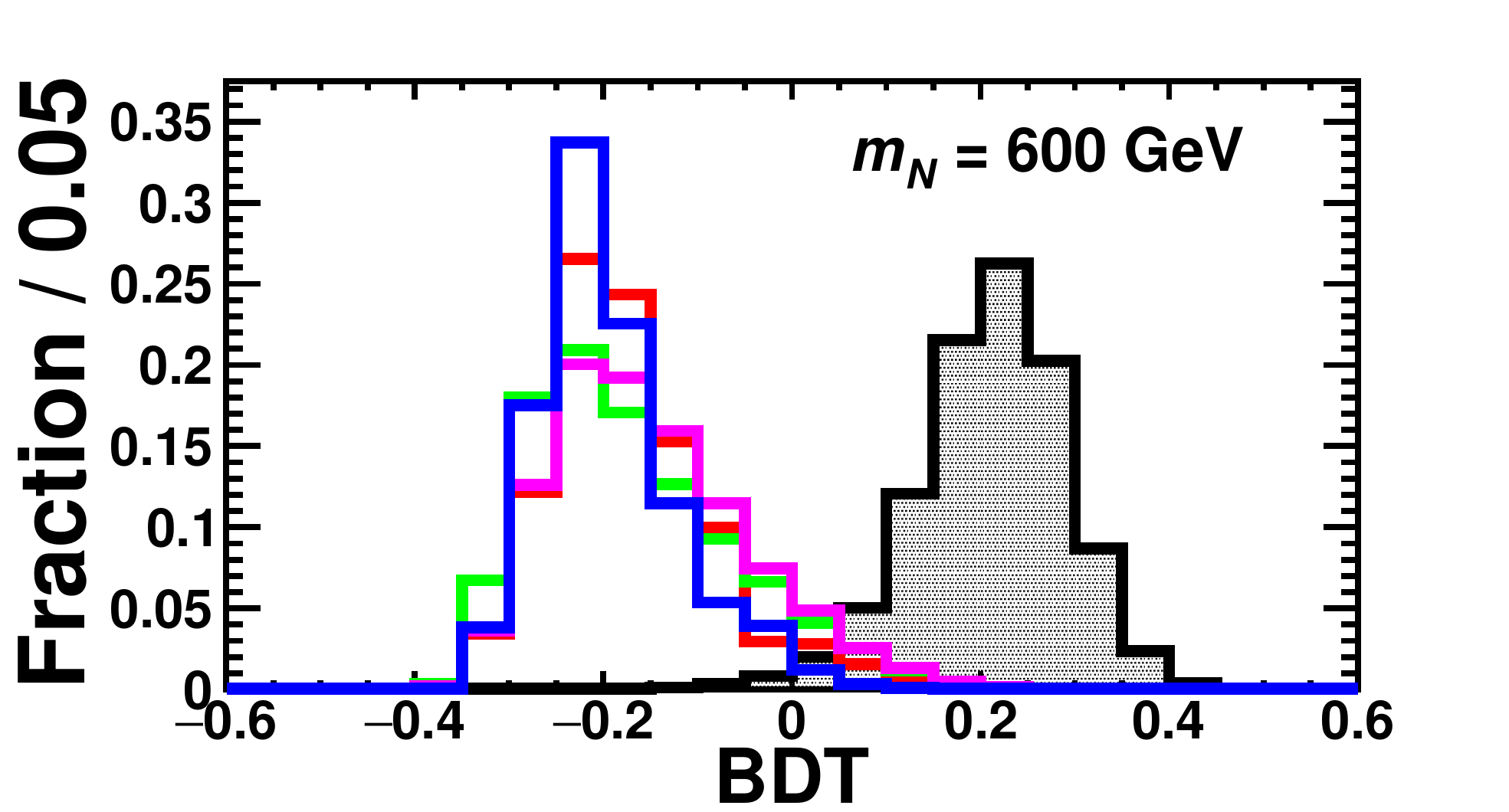}
}
\end{figure}
\vspace{-1.0cm}
\begin{figure}[H] 
\centering
\addtocounter{figure}{1}
\subfigure{
\includegraphics[width=4cm,height=3cm]{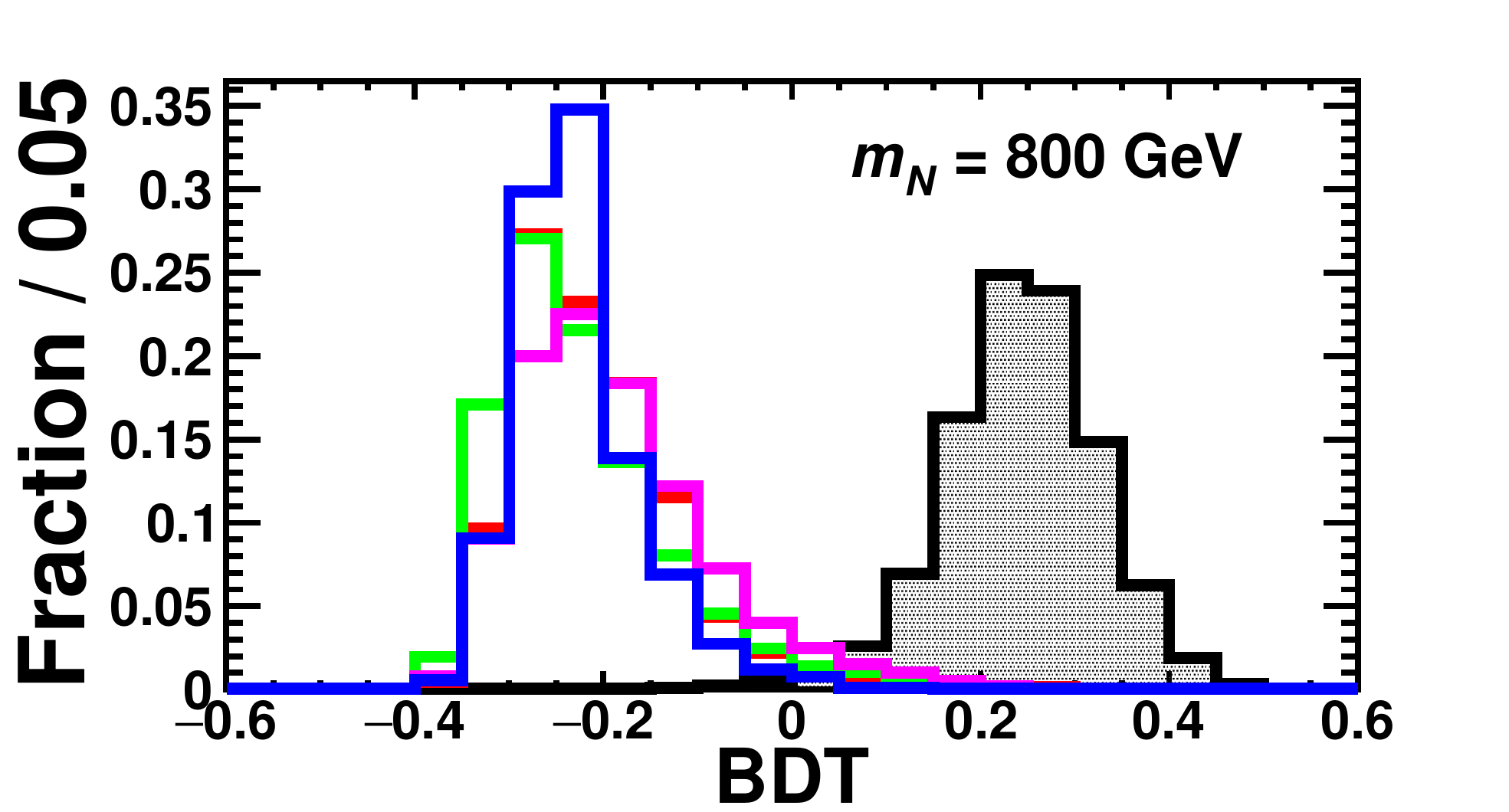}
\includegraphics[width=4cm,height=3cm]{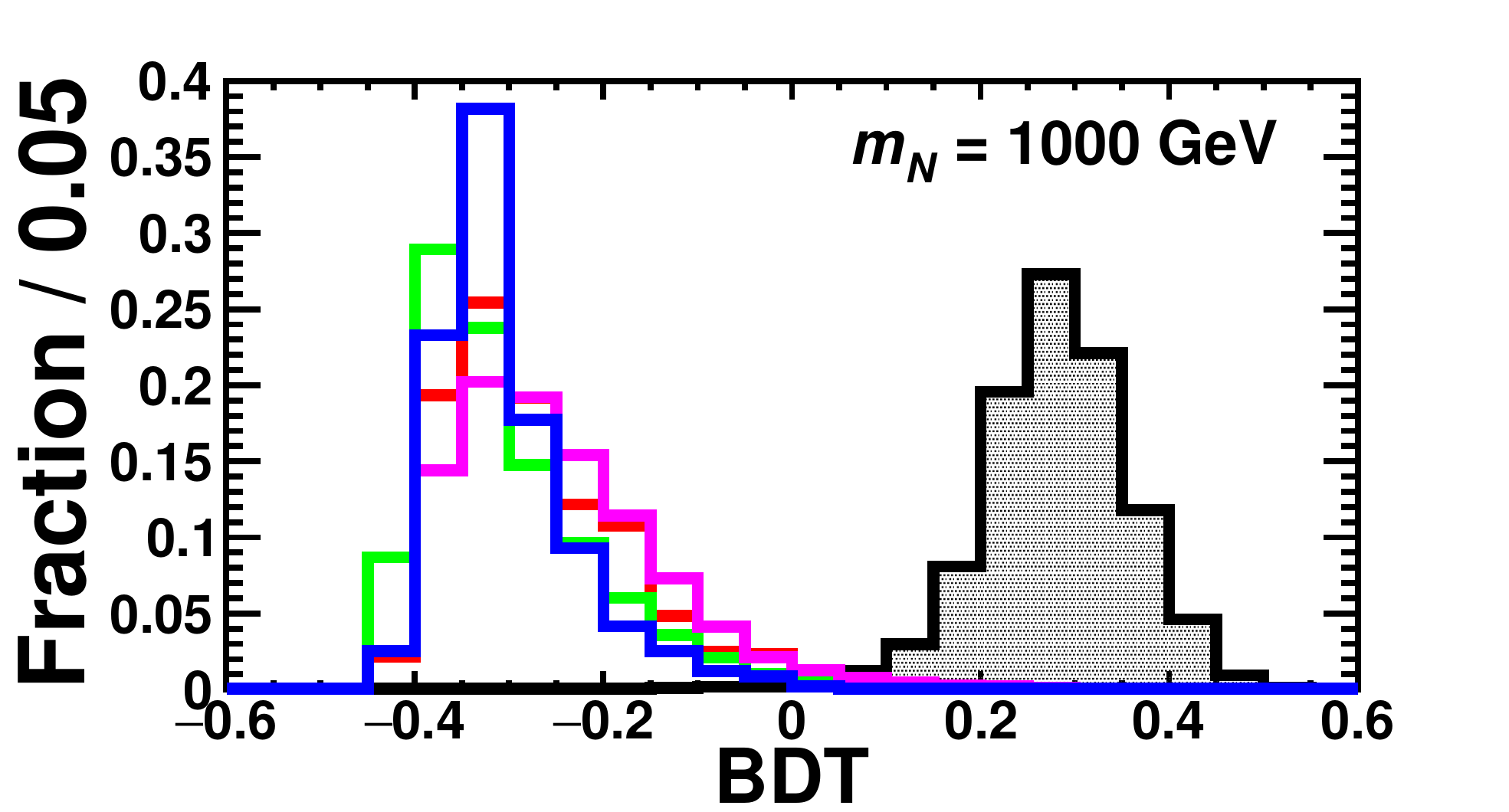}
}
\caption{The BDT distributions of the same processes as in Fig.~\ref{fig:BDTLHeC}, but at the FCC-eh.}
\label{fig:BDTFCC-eh}
\end{figure}

\section{Distributions of representative high-level observables}
\label{appendix:HLobs}

The following twenty-nine high-level observables are constructed and input to the TMVA package to perform the MVA-BDT analysis.
\begin{enumerate}[label*=(\alph*)]
\item 
The transverse momentum $p_{\rm T}$, the energy $E$, the pseudorapidity $\eta$ and the azimuthal angle $\phi$ of the final state particles:
$E(\mu)$, $E(j_{1})$, $E(j_{2})$, $E(j_{3})$,
$p_{\rm T}(\mu)$, $p_{\rm T}(j_{1})$, $p_{\rm T}(j_{2})$, $p_{\rm T}(j_{3})$,
$\eta(\mu)$, $\eta(j_{1})$, $\eta(j_{2})$, $\eta(j_{3})$,  
$\phi(\mu)$, $\phi(j_{1})$, $\phi(j_{2})$, $\phi(j_{3})$.
\item
The number of jets $N(j)$ and 
the magnitude and the azimuthal angle of the missing transverse momentum: 
$\met$, $\phi(\met)$.
\item 
 $p_{\rm T}$, $\eta$ and $\phi$ of the system of $(j_2 + j_3)$: $p_{\rm T}(j_{2}+j_{3})$, $\eta(j_{2}+j_{3})$, $\phi(j_{2}+j_{3})$.
\item 
The pseudorapidity difference $\Delta\eta$, the azimuthal angle difference $\Delta\phi$ and the angular distance difference $\Delta R=\sqrt{(\Delta\eta)^2+(\Delta\phi)^2}$ between the muon and jet(s):
$\Delta\eta(\mu,j_{1})$, $\Delta\eta(\mu,j_{2}+j_{3})$,
$\Delta\phi(\mu,j_{1})$, $\Delta\phi(\mu,j_{2}+j_{3})$,	
$\Delta R(\mu,j_{1})$, $\Delta R(\mu,j_{2}+j_{3})$.
\item 
The invariant mass $M$ of the system of $(\mu+j_{2}+j_{3})$: $M(\mu+j_{2}+j_{3})$.
\end{enumerate}	

\begin{figure}[H] 
\centering
	\subfigure{
		\includegraphics[width=4cm,height=3cm]{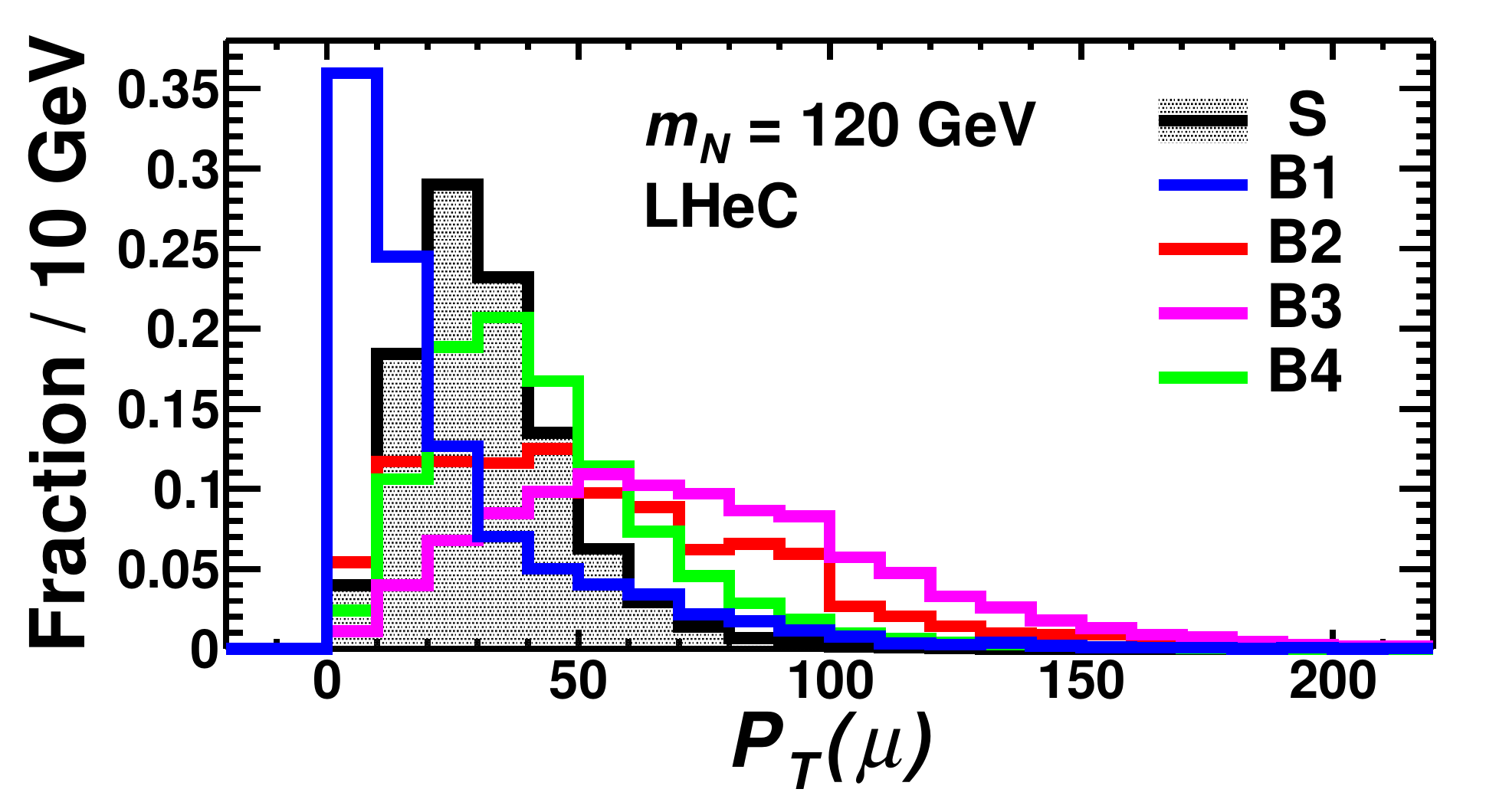}
		\includegraphics[width=4cm,height=3cm]{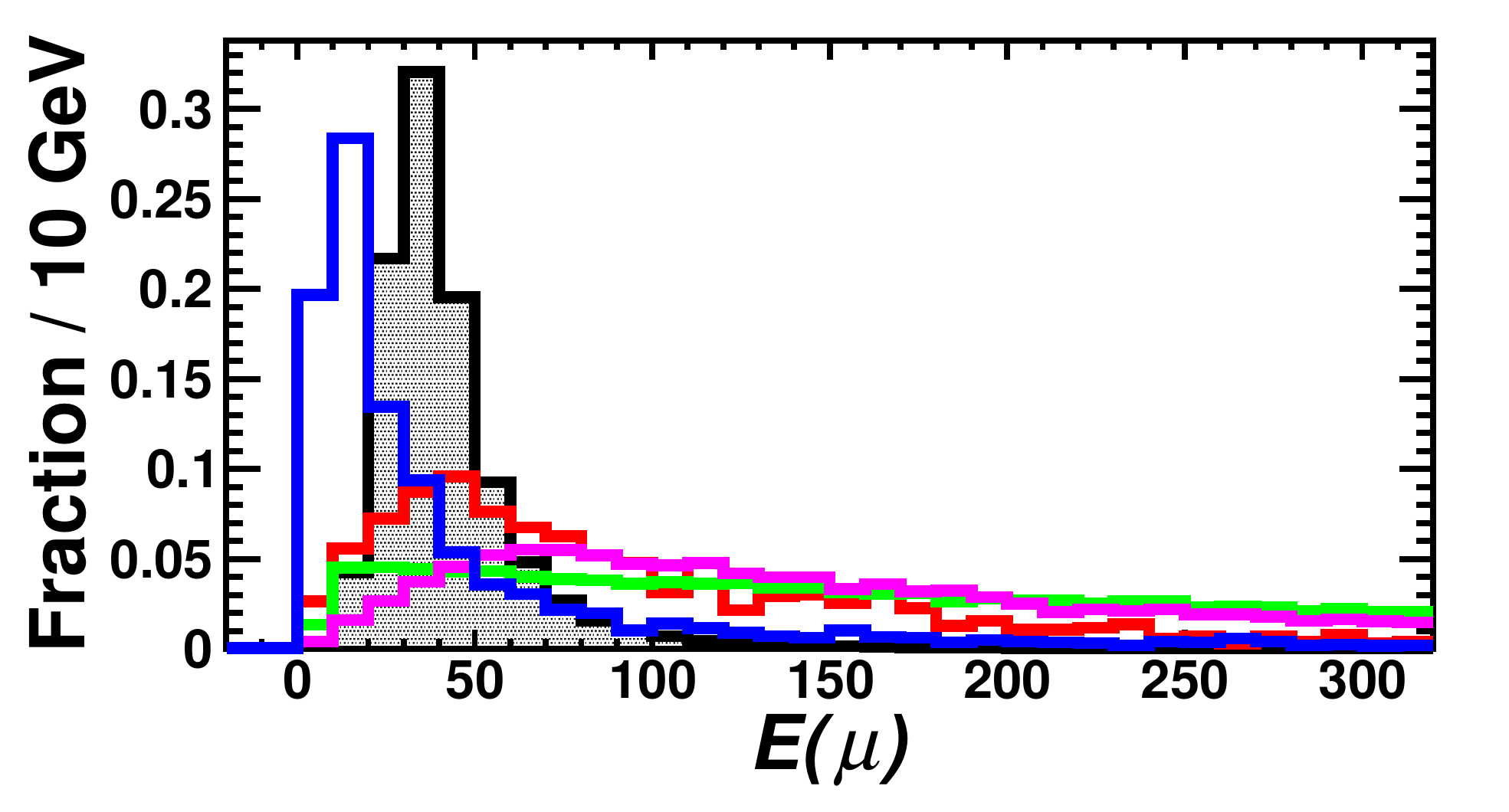}   
	}
\end{figure}
\addtocounter{figure}{-1}
\vspace{-1.0cm}
\begin{figure}[H] 
\centering
	\addtocounter{figure}{1}
	\subfigure{
			\includegraphics[width=4cm,height=3cm]{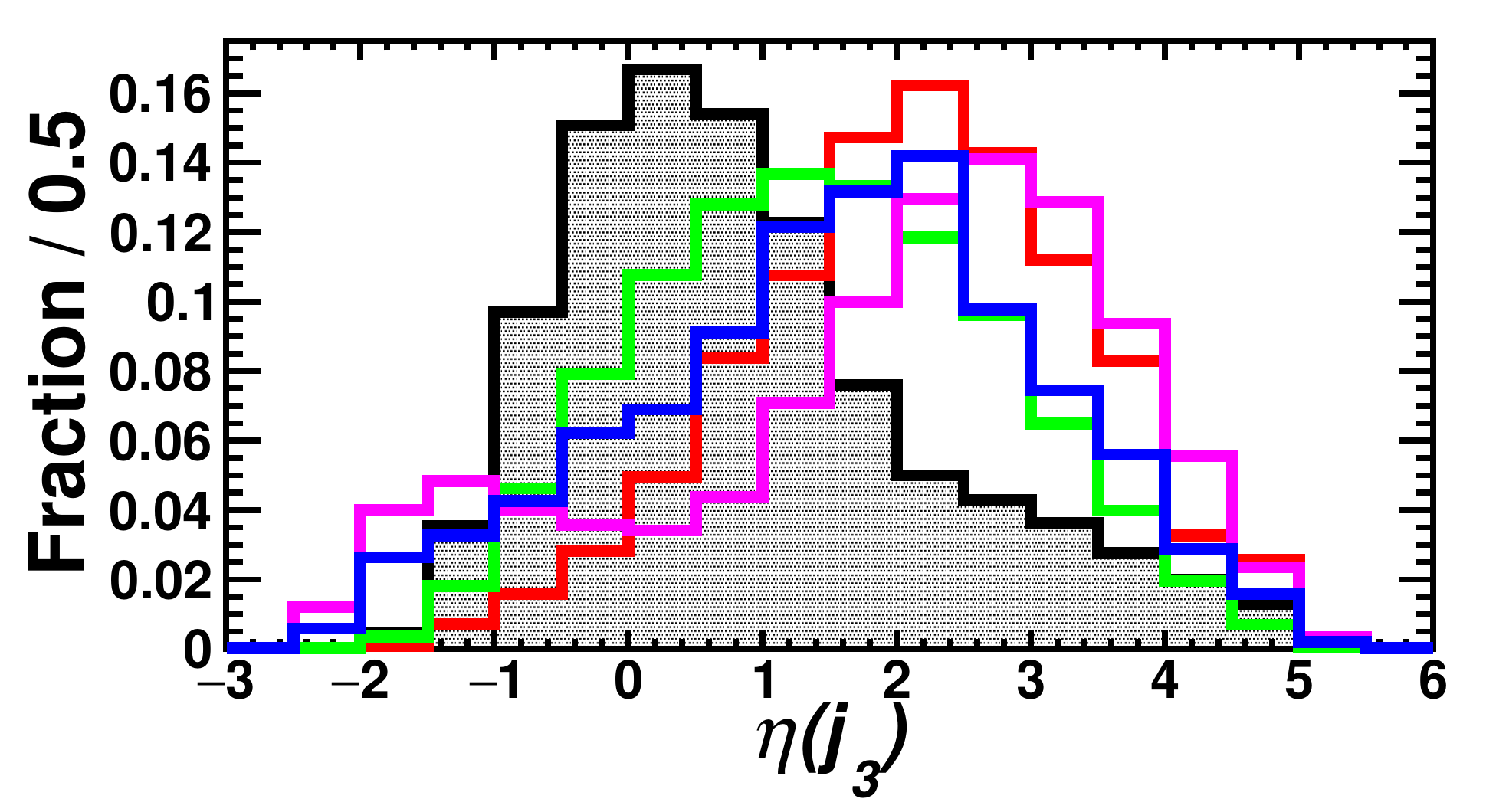}
		\includegraphics[width=4cm,height=3cm]{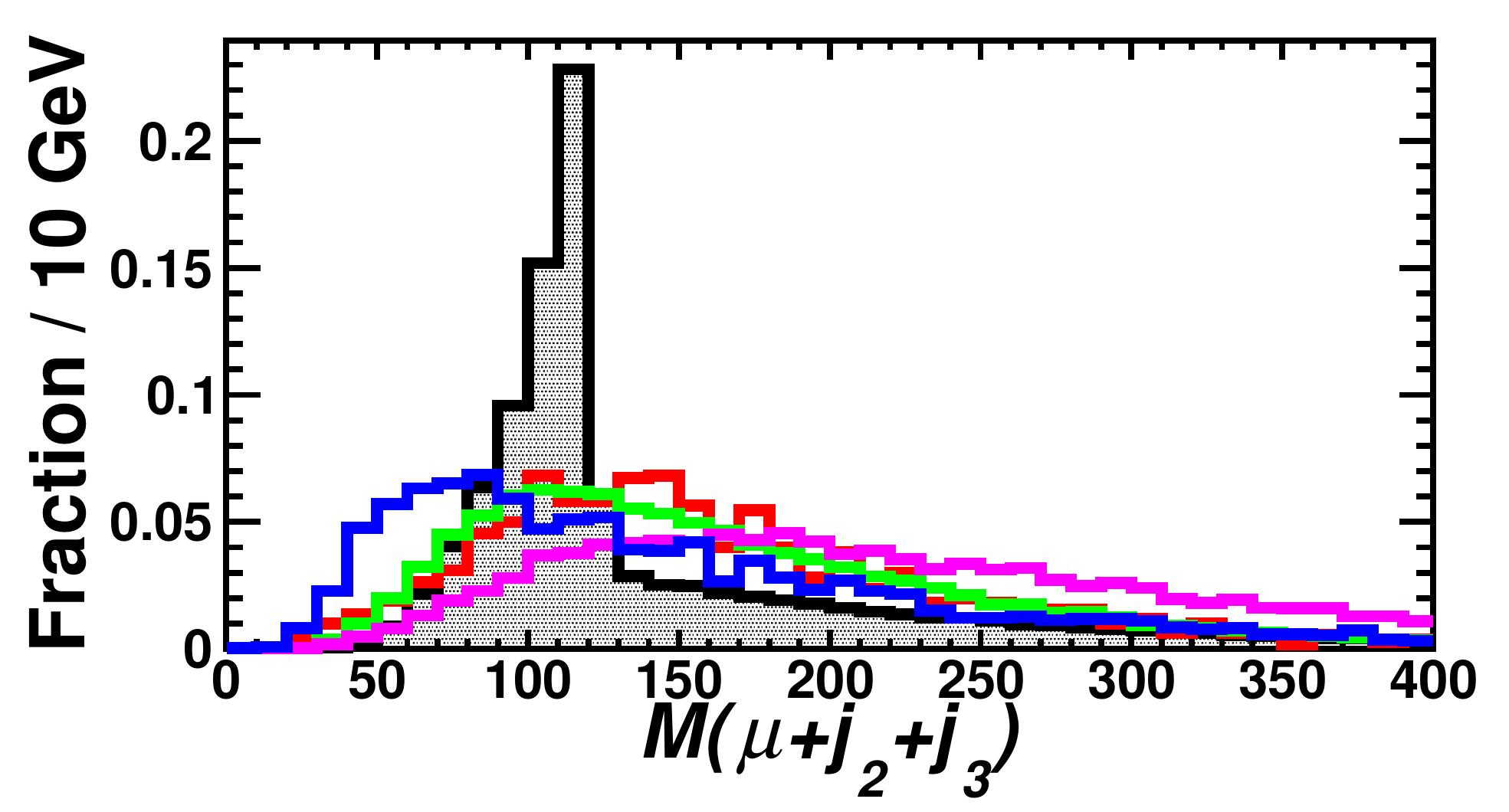} 
	}
\end{figure}
\vspace{-1.0cm}
\begin{figure}[H] 
\centering
	\addtocounter{figure}{-1}
	\subfigure{
		\includegraphics[width=4cm,height=3cm]{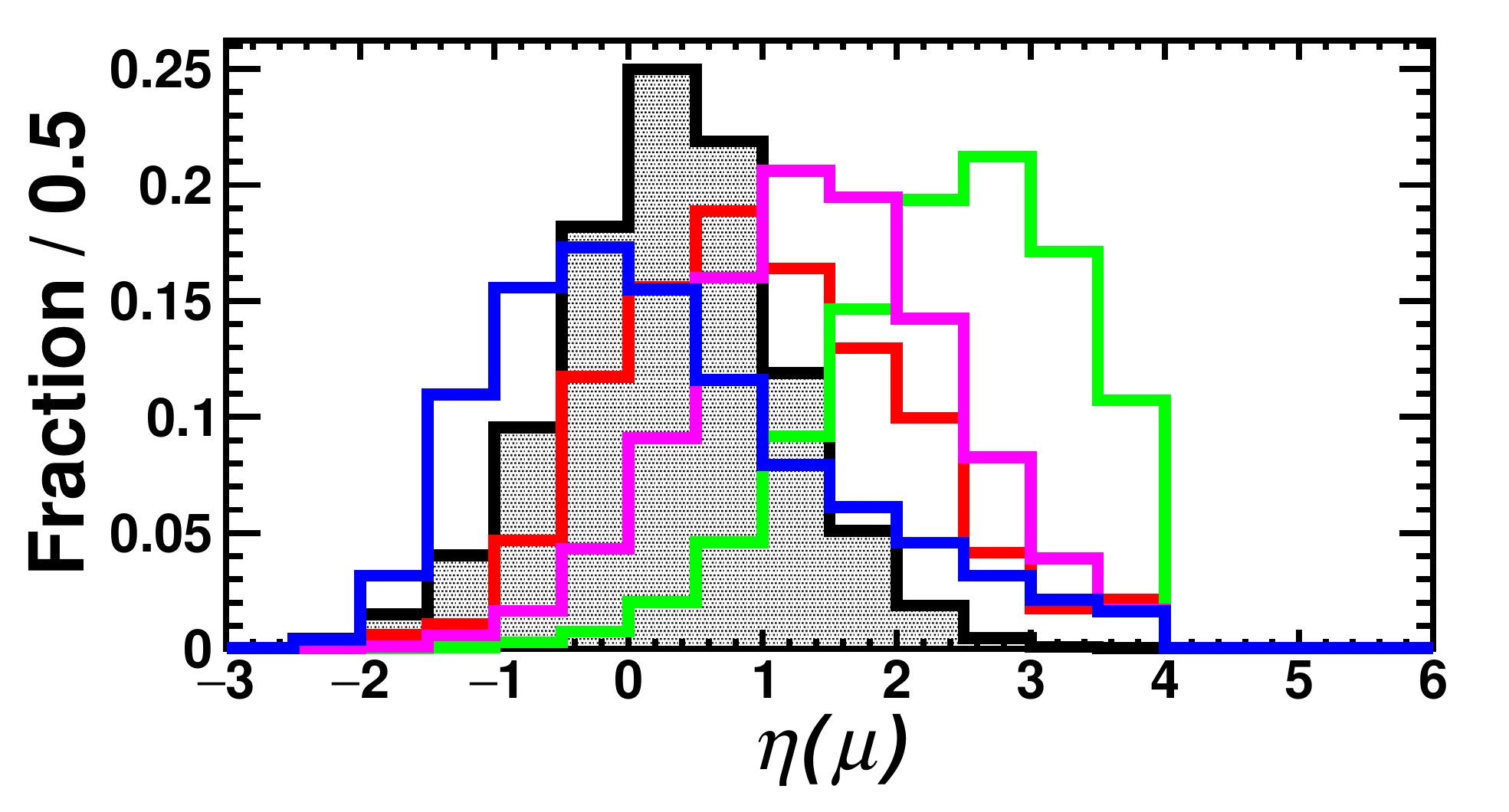}
		\includegraphics[width=4cm,height=3cm]{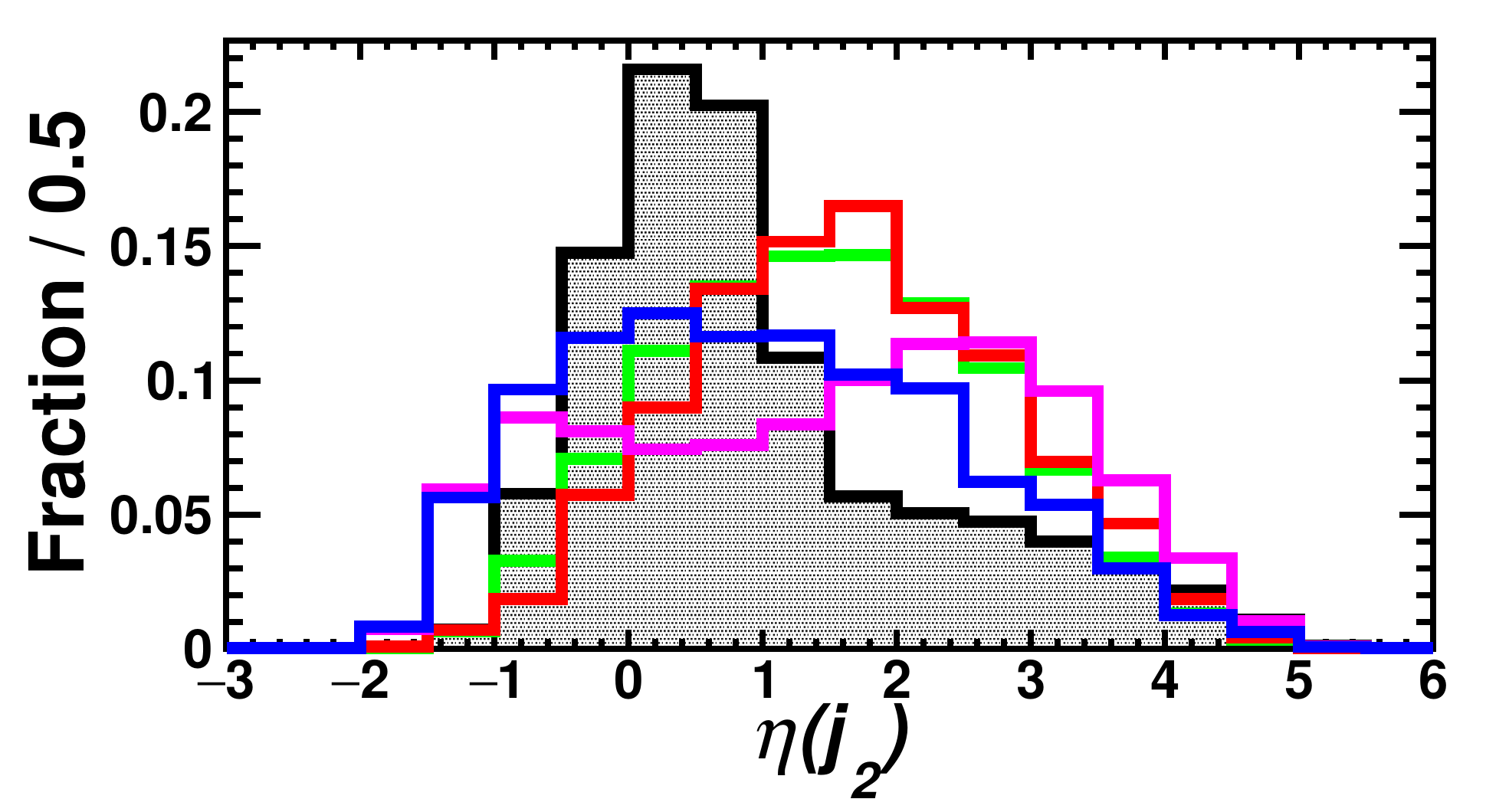} 
	}
\end{figure}
\vspace{-1.0cm}
\begin{figure}[H] 
\centering
	\addtocounter{figure}{1}
	\subfigure{
	\includegraphics[width=4cm,height=3cm]{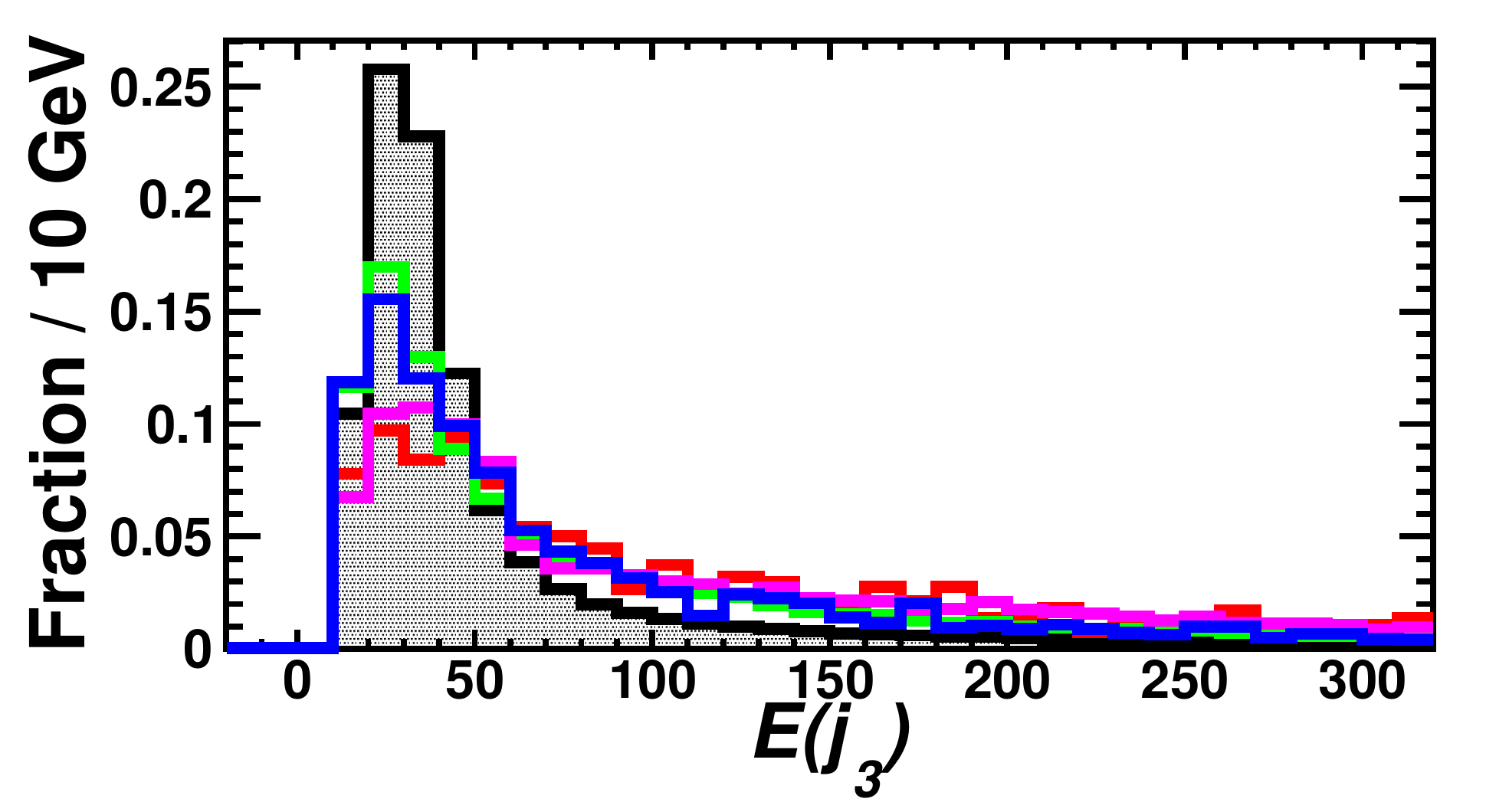} 
	\includegraphics[width=4cm,height=3cm]{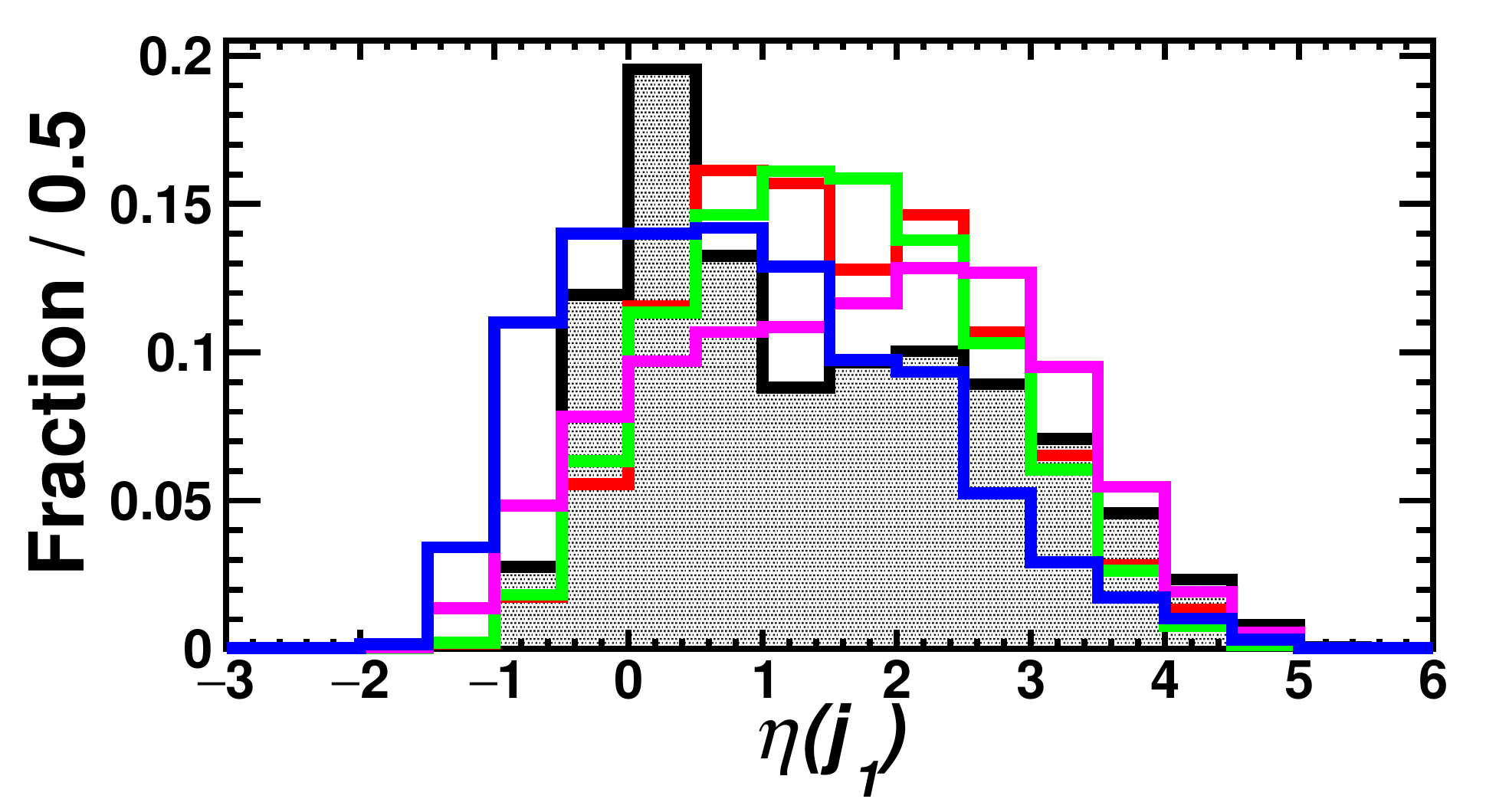} 
	}
\caption{
Distributions of representative high-level observables for the signal with $m_N = 120$ GeV (black, filled) and four background processes at the LHeC.
}
\label{fig:HLobsLHeC}
\end{figure}
\begin{figure}[H] 
\centering
	\subfigure{
		\includegraphics[width=4cm,height=3cm]{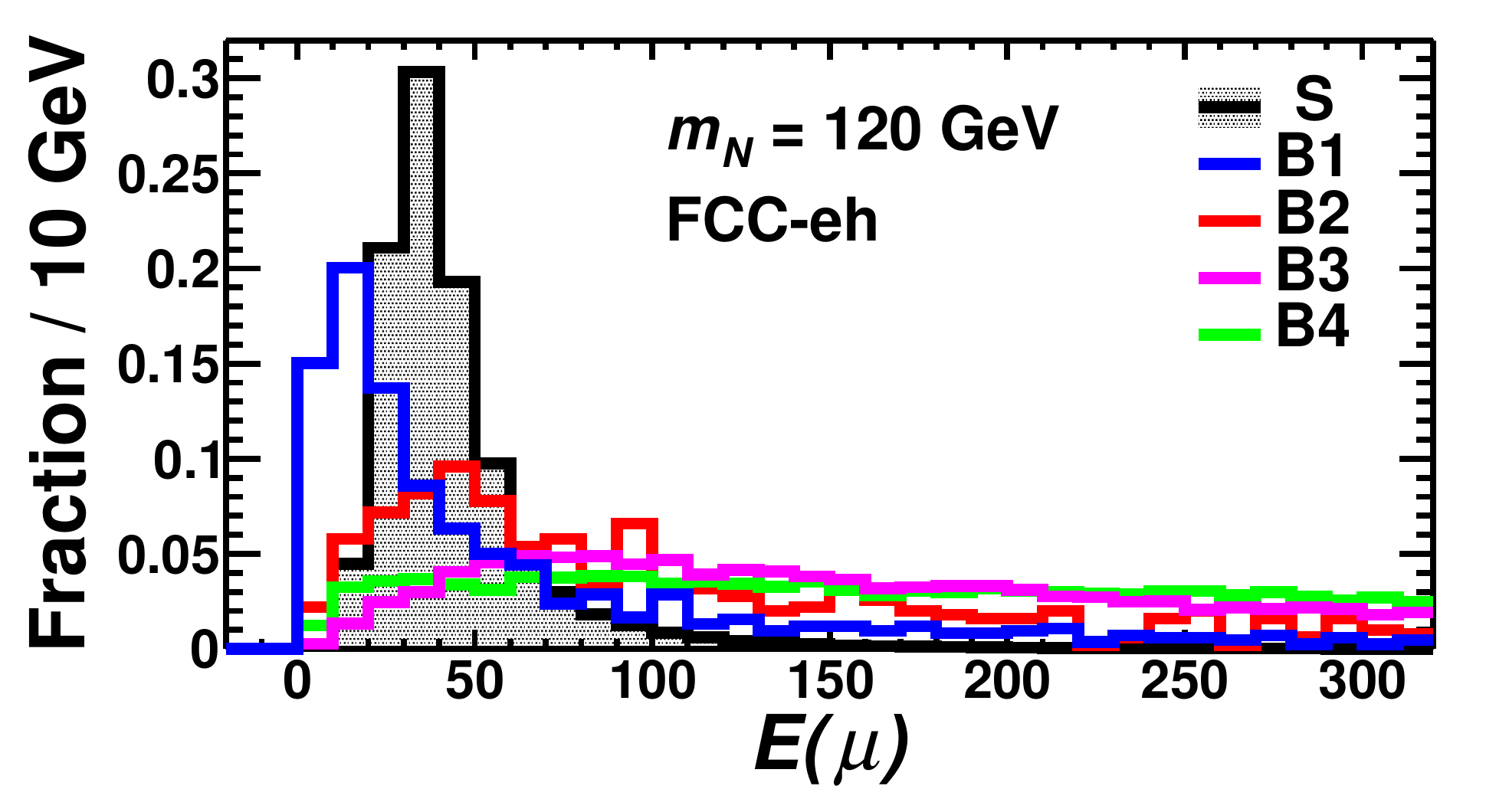}
		\includegraphics[width=4cm,height=3cm]{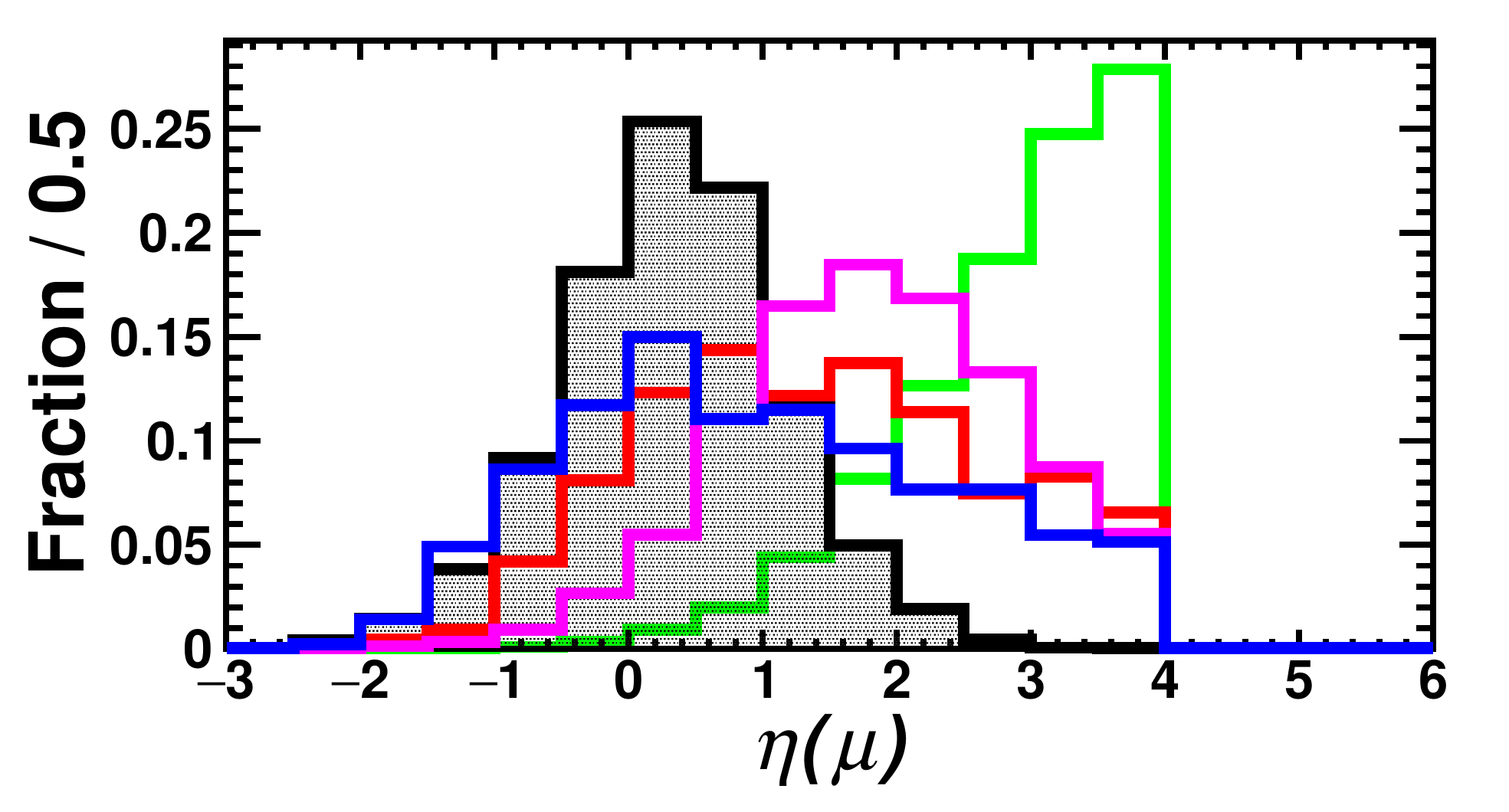}   
	}
\end{figure}
\addtocounter{figure}{-1}
\vspace{-1.0cm}
\begin{figure}[H] 
\centering
	\addtocounter{figure}{1}
	\subfigure{
	\includegraphics[width=4cm,height=3cm]{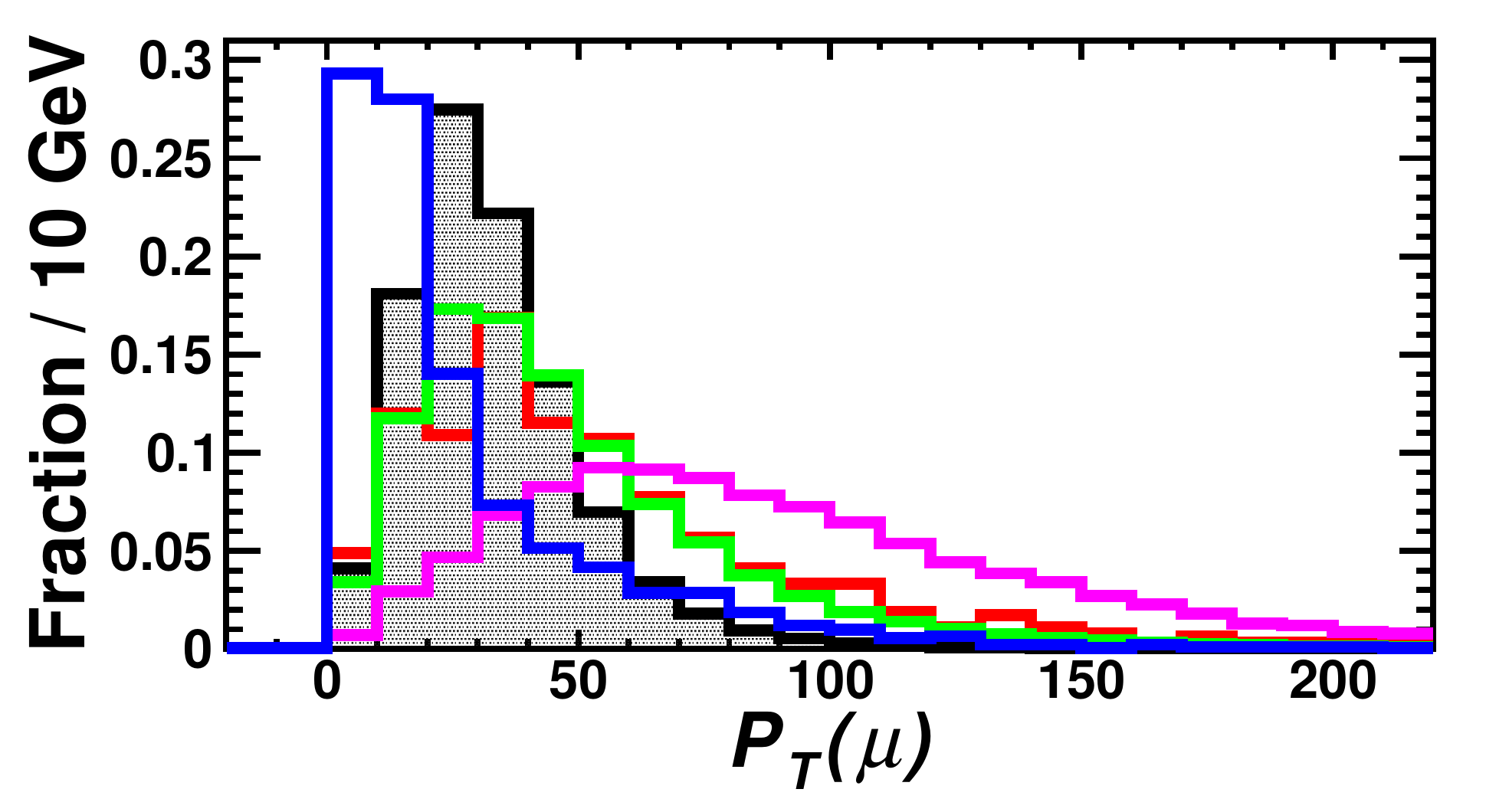}
	\includegraphics[width=4cm,height=3cm]{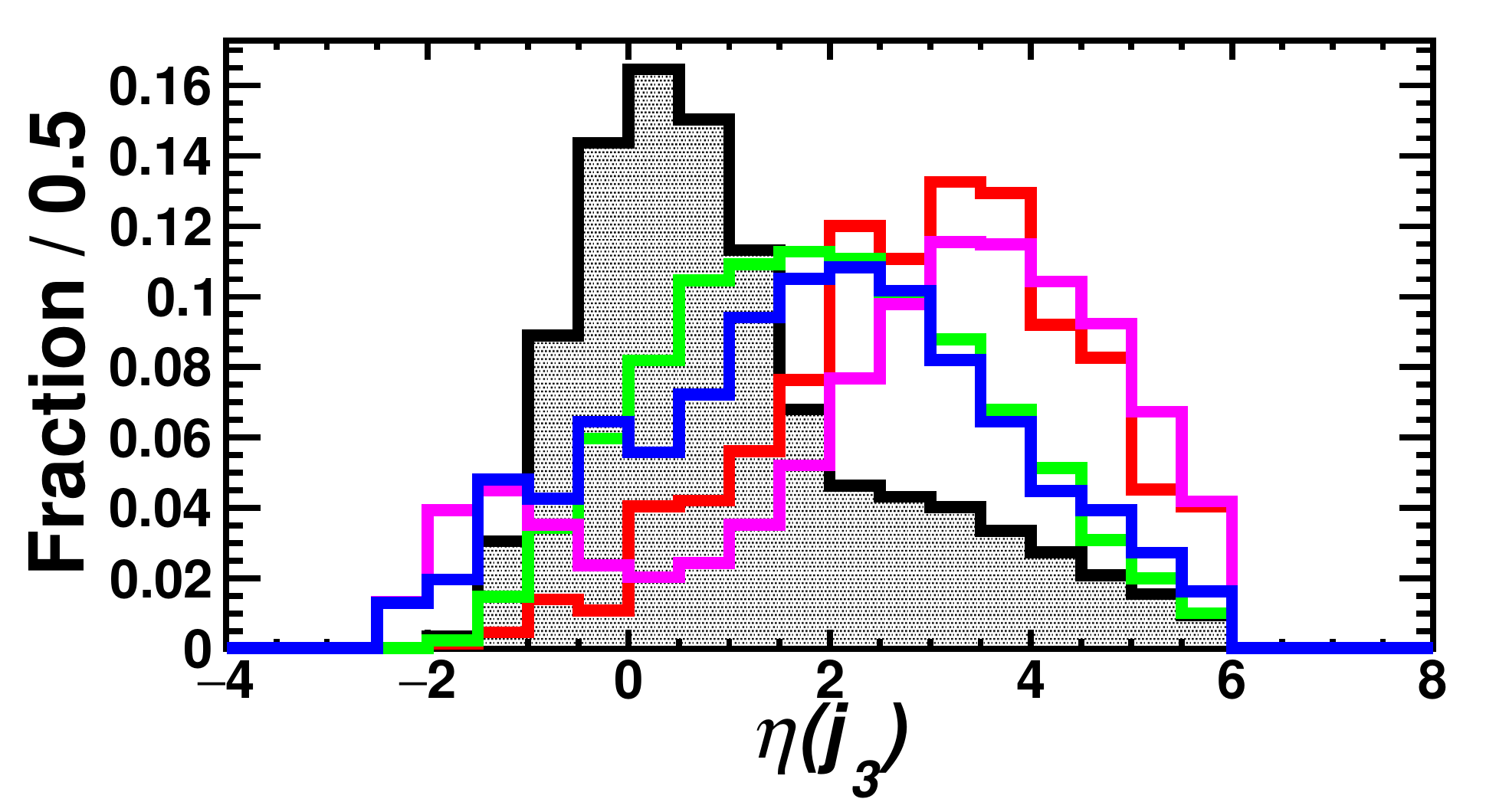} 
	}
\end{figure}
\vspace{-1.0cm}
\begin{figure}[H] 
\centering
	\addtocounter{figure}{-1}
	\subfigure{
	\includegraphics[width=4cm,height=3cm]{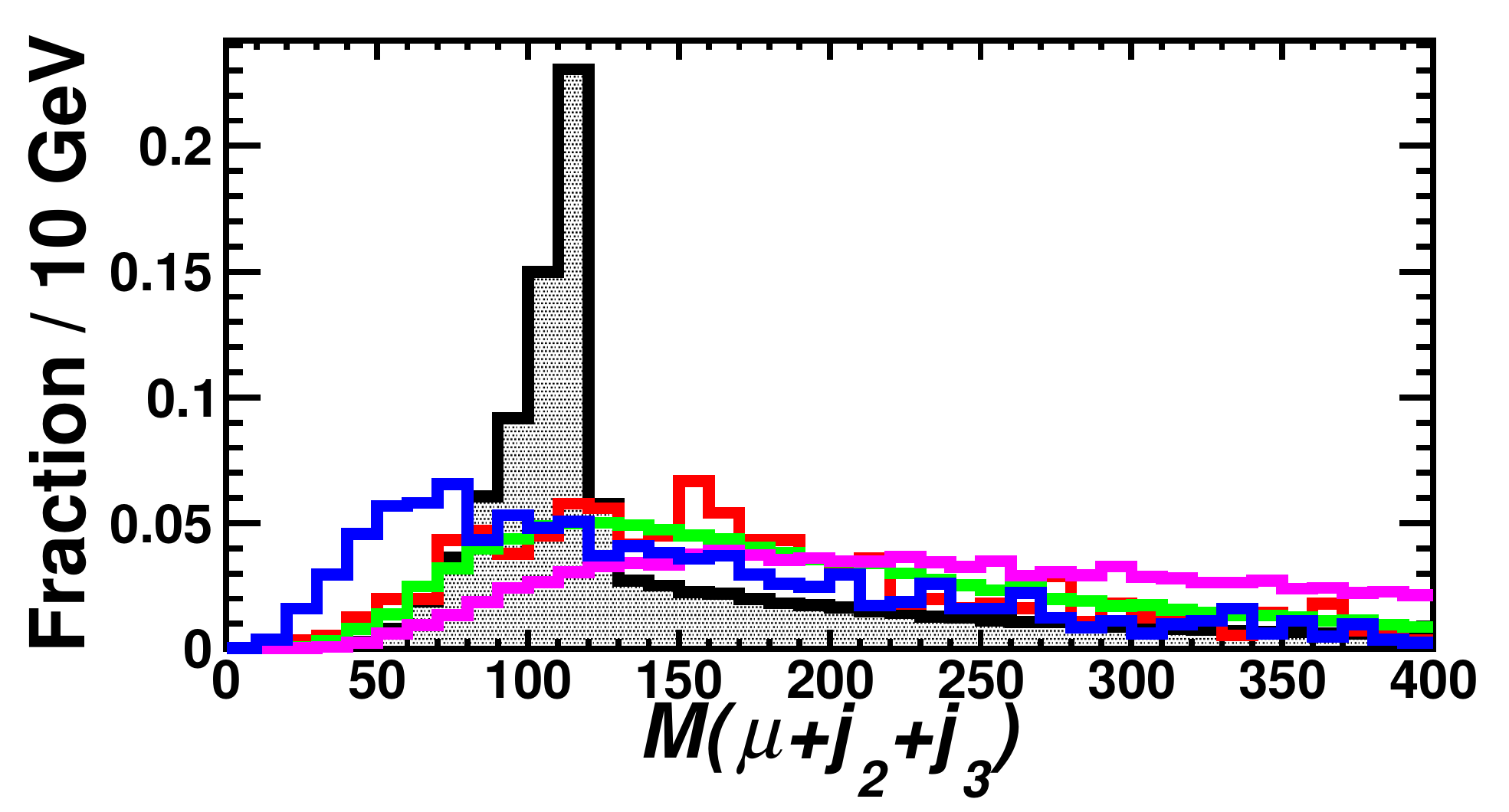}
	\includegraphics[width=4cm,height=3cm]{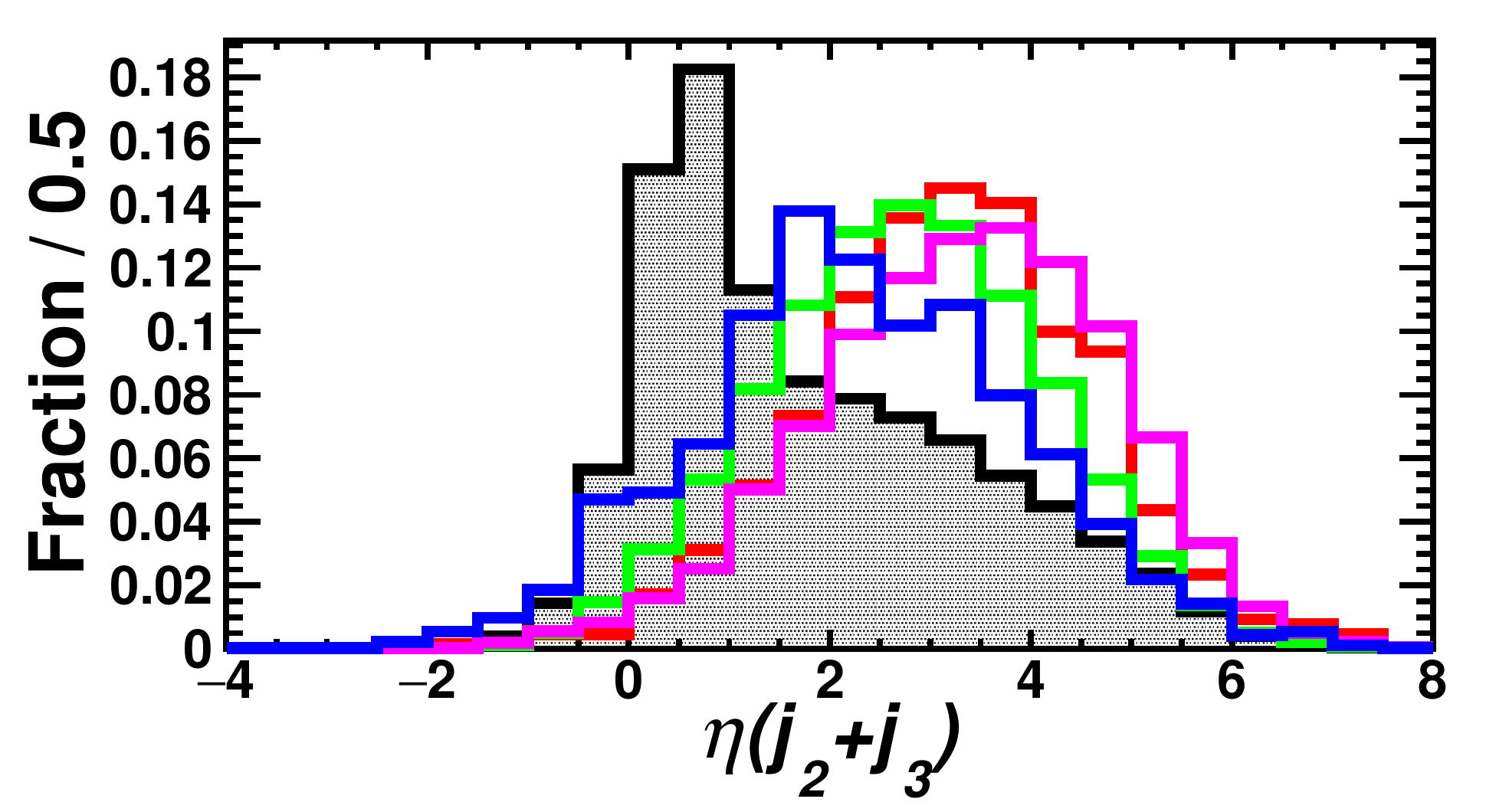} 
	}
\end{figure}
\vspace{-1.0cm}
\begin{figure}[H] 
\centering
	\addtocounter{figure}{1}
	\subfigure{
		\includegraphics[width=4cm,height=3cm]{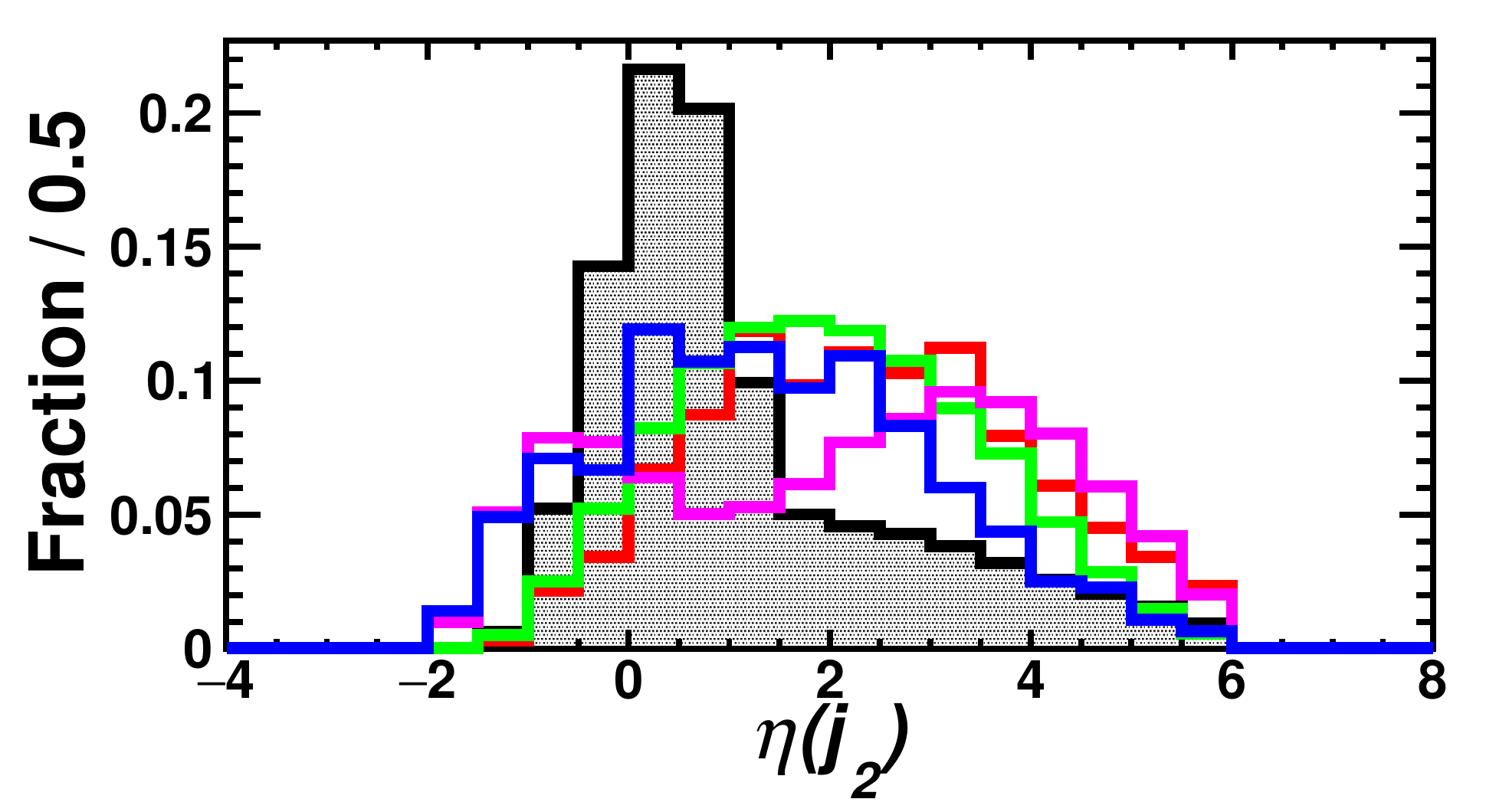} 
	\includegraphics[width=4cm,height=3cm]{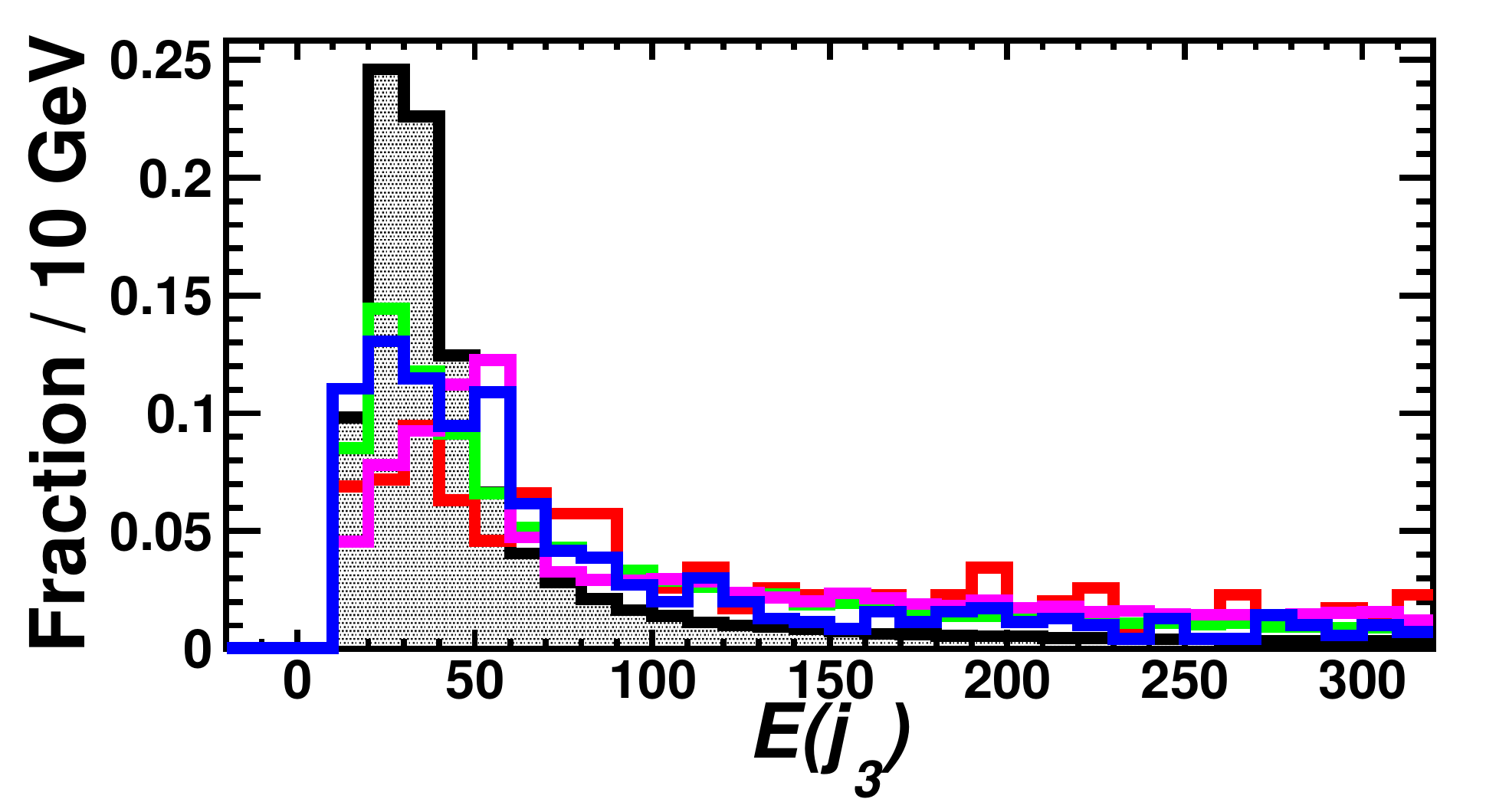} 
}
\caption{
The distributions of the same observables as in Fig.~\ref{fig:HLobsLHeC}, but at the FCC-eh.
}
\label{fig:HLobsFCCeh}
\end{figure}

\section{The Selection efficiency table}
\label{appendix:efficiencies}
In Table~\ref{tab:allEfficiencies}, we present selection efficiencies of pre-selection and BDT cuts for both signal and background processes at the LHeC and FCC-eh for representative heavy neutrino masses.
\begin{table*}[t]
\centering 
\begin{ruledtabular}
\begin{tabular}{cccccccc}
$m_N$ & collider & selection & signal & $\mu^+ \mu^- e^- jjj$ & $\mu^+ \mu^- \nu_e jjj$ & $\mu^+ \nu_\mu  e^- jjj$  & $\mu^+ \nu_\mu  \nu_e jjj$\\
\hline
\multirow{4}{*}{20 GeV} & \multirow{2}{*}{LHeC}    & pre-selection & $8.04\mltp10^{-2}$&$1.63\mltp10^{-3}$&$1.29\mltp10^{-4}$&$7.07\mltp10^{-4}$&$1.07\mltp10^{-2}$\\
& & BDT$>$0.070  &$8.53\mltp10^{-1}$&$4.57\mltp10^{-1}$&$7.61\mltp10^{-2}$&$2.00\mltp10^{-2}$&$1.59\mltp10^{-2}$\\
& \multirow{2}{*}{FCC-eh} &pre-selection&$1.20\mltp10^{-1}$&$6.10\mltp10^{-4}$&$1.22\mltp10^{-4}$&$7.86\mltp10^{-4}$&$4.71\mltp10^{-3}$\\
&                                            & BDT$>$0.112 &$8.25\mltp10^{-1}$&$3.13\mltp10^{-1}$&$4.54\mltp10^{-2}$&$1.04\mltp10^{-2}$&$7.52\mltp10^{-3}$\\
\hline
\multirow{4}{*}{40 GeV} & \multirow{2}{*}{LHeC}    & pre-selection & $2.40\mltp10^{-1}$&$1.63\mltp10^{-3}$&$1.29\mltp10^{-4}$&$7.07\mltp10^{-4}$&$1.07\mltp10^{-2}$\\
& & BDT$>$0.125  &$4.76\mltp10^{-1}$&$6.99\mltp10^{-2}$&$1.62\mltp10^{-2}$&$2.69\mltp10^{-3}$&$5.30\mltp10^{-3}$\\
& \multirow{2}{*}{FCC-eh} &pre-selection&$2.77\mltp10^{-1}$&$6.10\mltp10^{-4}$&$1.22\mltp10^{-4}$&$7.86\mltp10^{-4}$&$4.71\mltp10^{-3}$\\
 &                                            & BDT$>$0.109 &$6.55\mltp10^{-1}$&$3.07\mltp10^{-1}$&$4.13\mltp10^{-2}$&$8.75\mltp10^{-3}$&$7.64\mltp10^{-3}$\\
\hline
\multirow{4}{*}{60 GeV} & \multirow{2}{*}{LHeC}    & pre-selection & $3.71\mltp10^{-1}$&$1.63\mltp10^{-3}$&$1.29\mltp10^{-4}$&$7.07\mltp10^{-4}$&$1.07\mltp10^{-2}$\\
& & BDT$>$0.116  &$5.43\mltp10^{-1}$&$8.38\mltp10^{-2}$&$2.88\mltp10^{-2}$&$5.37\mltp10^{-3}$&$1.19\mltp10^{-2}$\\
& \multirow{2}{*}{FCC-eh} &pre-selection&$3.91\mltp10^{-1}$&$6.10\mltp10^{-4}$&$1.22\mltp10^{-4}$&$7.86\mltp10^{-4}$&$4.71\mltp10^{-3}$\\
&                                            & BDT$>$0.111 &$7.73\mltp10^{-1}$&$3.01\mltp10^{-1}$&$4.40\mltp10^{-2}$&$1.07\mltp10^{-2}$&$1.39\mltp10^{-2}$\\
\hline
\multirow{4}{*}{120 GeV} & \multirow{2}{*}{LHeC}    & pre-selection & $5.66\mltp10^{-1}$&$1.20\mltp10^{-3}$&$1.08\mltp10^{-4}$&$5.53\mltp10^{-4}$&$9.07\mltp10^{-3}$\\
& & BDT$>$0.067  &$7.71\mltp10^{-1}$&$1.21\mltp10^{-1}$&$8.36\mltp10^{-2}$&$2.29\mltp10^{-2}$&$4.74\mltp10^{-2}$\\
& \multirow{2}{*}{FCC-eh} &pre-selection&$5.45\mltp10^{-1}$&$4.53\mltp10^{-4}$&$1.08\mltp10^{-4}$&$6.64\mltp10^{-4}$&$4.22\mltp10^{-3}$\\
&                                            & BDT$>$0.094 &$7.95\mltp10^{-1}$&$1.72\mltp10^{-1}$&$7.49\mltp10^{-2}$&$1.63\mltp10^{-2}$&$2.56\mltp10^{-2}$\\
\hline
\multirow{4}{*}{200 GeV} & \multirow{2}{*}{LHeC}    & pre-selection & $6.00\mltp10^{-1}$&$1.20\mltp10^{-3}$&$1.08\mltp10^{-4}$&$5.53\mltp10^{-4}$&$9.07\mltp10^{-3}$\\
& & BDT$>$0.087  &$6.84\mltp10^{-1}$&$4.10\mltp10^{-2}$&$5.56\mltp10^{-2}$&$3.01\mltp10^{-2}$&$3.60\mltp10^{-2}$\\
& \multirow{2}{*}{FCC-eh} &pre-selection&$5.57\mltp10^{-1}$&$4.53\mltp10^{-4}$&$1.08\mltp10^{-4}$&$6.64\mltp10^{-4}$&$4.22\mltp10^{-3}$\\
&                                            & BDT$>$0.102 &$6.59\mltp10^{-1}$&$5.91\mltp10^{-2}$&$4.37\mltp10^{-2}$&$2.74\mltp10^{-2}$&$2.74\mltp10^{-2}$\\
\hline
\multirow{4}{*}{400 GeV} & \multirow{2}{*}{LHeC}    & pre-selection & $4.33\mltp10^{-1}$&$1.20\mltp10^{-3}$&$1.08\mltp10^{-4}$&$5.53\mltp10^{-4}$&$9.07\mltp10^{-3}$\\
& & BDT$>$0.154  &$6.98\mltp10^{-1}$&$2.37\mltp10^{-3}$&$7.05\mltp10^{-3}$&$9.45\mltp10^{-3}$&$9.63\mltp10^{-3}$\\
& \multirow{2}{*}{FCC-eh} &pre-selection&$4.67\mltp10^{-1}$&$4.53\mltp10^{-4}$&$1.08\mltp10^{-4}$&$6.64\mltp10^{-4}$&$4.22\mltp10^{-3}$\\
&                                            & BDT$>$0.140 &$6.64\mltp10^{-1}$&$3.28\mltp10^{-3}$&$1.40\mltp10^{-2}$&$1.26\mltp10^{-2}$&$1.07\mltp10^{-2}$\\
\hline
\multirow{4}{*}{600 GeV} & \multirow{2}{*}{LHeC}    & pre-selection & $2.61\mltp10^{-1}$&$1.20\mltp10^{-3}$&$1.08\mltp10^{-4}$&$5.53\mltp10^{-4}$&$9.07\mltp10^{-3}$\\
& & BDT$>$0.247 &$5.91\mltp10^{-1}$&$-$&$1.76\mltp10^{-3}$&$6.61\mltp10^{-4}$&$5.33\mltp10^{-4}$\\
& \multirow{2}{*}{FCC-eh} &pre-selection&$3.66\mltp10^{-1}$&$4.53\mltp10^{-4}$&$1.08\mltp10^{-4}$&$6.64\mltp10^{-4}$&$4.22\mltp10^{-3}$\\
&                                            & BDT$>$0.182 &$6.66\mltp10^{-1}$&$-$&$-$&$3.43\mltp10^{-3}$&$1.78\mltp10^{-3}$\\
\hline
\multirow{4}{*}{800 GeV} & \multirow{2}{*}{LHeC}    & pre-selection & $1.55\mltp10^{-1}$&$1.20\mltp10^{-3}$&$1.08\mltp10^{-4}$&$5.53\mltp10^{-4}$&$9.07\mltp10^{-3}$\\
& & BDT$>$0.210 &$9.27\mltp10^{-1}$&$-$&$-$&$1.12\mltp10^{-3}$&$6.25\mltp10^{-4}$\\
& \multirow{2}{*}{FCC-eh} &pre-selection&$2.57\mltp10^{-1}$&$4.53\mltp10^{-4}$&$1.08\mltp10^{-4}$&$6.64\mltp10^{-4}$&$4.22\mltp10^{-3}$\\
&                                            & BDT$>$0.211 &$6.71\mltp10^{-1}$&$-$&$1.56\mltp10^{-3}$&$2.08\mltp10^{-3}$&$1.26\mltp10^{-3}$\\
\hline
\multirow{4}{*}{1000 GeV} & \multirow{2}{*}{LHeC}    & pre-selection & $7.92\mltp10^{-2}$&$1.20\mltp10^{-3}$&$1.08\mltp10^{-4}$&$5.53\mltp10^{-4}$&$9.07\mltp10^{-3}$\\
& & BDT$>$0.238 &$9.63\mltp10^{-1}$&$-$&$-$&$3.30\mltp10^{-4}$&$5.51\mltp10^{-5}$\\
& \multirow{2}{*}{FCC-eh} &pre-selection&$1.60\mltp10^{-1}$&$4.53\mltp10^{-4}$&$1.08\mltp10^{-4}$&$6.64\mltp10^{-4}$&$4.22\mltp10^{-3}$\\
&                                            & BDT$>$0.227 &$7.73\mltp10^{-1}$&$-$&$1.56\mltp10^{-3}$&$2.57\mltp10^{-3}$&$1.15\mltp10^{-3}$\\
                                         
\end{tabular}
\end{ruledtabular}
\caption{
Selection efficiencies of pre-selection and BDT cuts for both signal and background processes at the LHeC and FCC-eh for representative heavy neutrino masses,
where ``$-$" means the number of events can be reduced to be negligible.
}
\label{tab:allEfficiencies}
\end{table*}


\begin{acknowledgments}
\noindent
We thank Lingxiao Bai, Marco Drewes, Filmon Andom Ghebretinsae, Ying-nan Mao, Minglun Tian and Zeren Simon Wang for helpful discussions. 
H.G. and K.W. are supported by the National Natural Science Foundation of China under grant no.~11905162, 
the Excellent Young Talents Program of the Wuhan University of Technology under grant no.~40122102, and the research program of the Wuhan University of Technology under grant no.~2020IB024.
The simulation and analysis work of this paper was completed with the computational cluster provided by the Theoretical Physics Group at the Department of Physics, School of Sciences, Wuhan University of Technology.
\end{acknowledgments}


\bibliography{Refs}

\bibliographystyle{h-physrev5}

\end{document}